\documentclass[]{spie}  

\usepackage{graphicx,amsmath,amssymb,subcaption,subfig}
\title{Optimized focal and pupil plane masks for vortex coronagraphs on telescopes with obstructed apertures} 

\author{Garreth J. Ruane\supit{a,b}, Olivier Absil\supit{b}, Elsa Huby\supit{b}, Dimitri Mawet\supit{c}, Christian Delacroix\supit{d}, Brunella Carlomagno\supit{b}, Pierre Piron\supit{b}, and Grover A. Swartzlander, Jr.\supit{a}
\skiplinehalf
\supit{a}Chester F. Carlson Center for Imaging Science, Rochester Institute of Technology, \\54 Lomb Memorial Drive, Rochester, NY 14623, USA; \\
\supit{b}D\'{e}partement d'Astrophysique, G\'{e}ophysique et Oc\'{e}anographie, Universit\'{e} de Li\`{e}ge, \\All\'{e}e du Six Ao\^{u}t 17, 4000 Li\`{e}ge, Belgium;\\
\supit{c}California Institute of Technology, 1200 E. California Blvd., Pasadena, CA 91125, USA;\\
\supit{d}CRAL, Observatoire de Lyon, CNRS UMR 5574, Universit\'{e} Lyon 1, 9 avenue Charles Andr\'{e}e, 69230 Saint-Genis Laval, France
}
\authorinfo{Further author information: send correspondence to gjr8334@rit.edu}

\begin{document} 
\maketitle 
\begin{abstract}
We present methods for optimizing pupil and focal plane optical elements that improve the performance of vortex coronagraphs on telescopes with obstructed or segmented apertures. Phase-only and complex masks are designed for the entrance pupil, focal plane, and the plane of the Lyot stop. Optimal masks are obtained using both analytical and numerical methods. The latter makes use of an iterative error reduction algorithm to calculate ``correcting'' optics that mitigate unwanted diffraction from aperture obstructions. We analyze the achieved performance in terms of starlight suppression, contrast, off-axis image quality, and chromatic dependence. Manufacturing considerations and sensitivity to aberrations are also discussed. This work provides a path to joint optimization of multiple coronagraph planes to maximize sensitivity to exoplanets and other faint companions. 
\end{abstract}


\section{INTRODUCTION}
Direct imaging and characterization of exoplanets requires an optical system that can selectively suppress light from a star that would otherwise inhibit detection of dim companions and their spectroscopic signatures. Such observations have been made increasingly possible by dedicated coronagraphic high-contrast imaging instruments with extreme adaptive optics, such as GPI \cite{GPI2006}, VLT/SPHERE \cite{SPHERE2008}, and Subaru/HiCIAO \cite{Hodapp2008}. 
  
Many elegant optical designs for coronagraphs are available, the most common of which fall into two categories: focal-plane \cite{Lyot1939,Roddier1997,Rouan2000,Kuchner2002,Soummer2003a,Mawet2005,Foo2005,Trauger2007,Serabyn2010} and pupil-plane \cite{Kasdin2003,Codona2004,Kenworthy2007} coronagraphs. The former make use of a focal plane mask and downstream ``Lyot stop" (LS) to block on-axis starlight from reaching the detector. The latter have a single amplitude or phase mask in the pupil of an imaging system which alters the point spread function (PSF) such that a dark hole appears in the image of the star where faint exoplanets may be detected. More advanced designs may incorporate multiple pupil plane, focal plane, and out-of-plane pupil remapping optics \cite{Guyon2003,Guyon2005,Soummer2003b}. 

Many of the coronagraph architectures mentioned above function well with an unobstructed, circular pupil. However, the performance is often severely degraded on telescopes with non-circular pupils and/or in the presence of aperture obstructions, such as a secondary mirror, spider support structures, and gaps between mirror segments. In such cases, advanced optical designs are required for high-performance coronagraphy. The optical system may be optimized with several performance aspects in mind including contrast, throughput, chromatic dependence, and sensitivity to aberrations \cite{Soummer2005,Carlotti2011,Carlotti2013,Pueyo2013,Guyon2014}.

This work introduces optical elements designed to compensate for unwanted diffraction owing to pupil obstructions. We consider the example of the vortex coronagraph (VC), which in its conventional design makes use of a focal plane phase mask and binary amplitude LS \cite{Mawet2005,Foo2005}. The VC has a number of advantages, including a small inner working angle (IWA) and intrinsic achromaticity, but it is very sensitive to the pupil shape. To tailor the VC for a complicated telescope aperture, we introduce phase-only and complex field correctors for the entrance pupil, focal plane, and/or Lyot plane and discuss the resulting performance gains. 

\begin{figure}[t!]
\begin{center}
\includegraphics{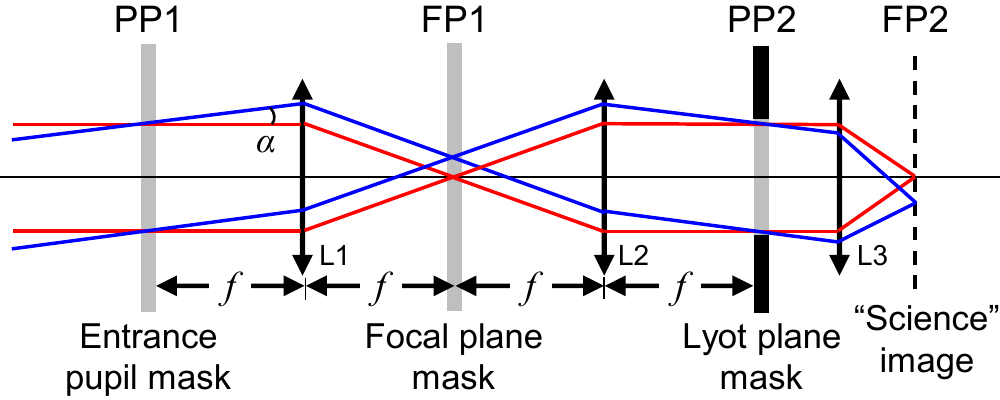}
\end{center}
\caption{ \label{fig:generalschematic} 
A generalized coronagraph with masks at the entrance pupil, focal plane, and Lyot plane.}
\end{figure} 

\begin{figure}[b!]
\begin{center}
\includegraphics{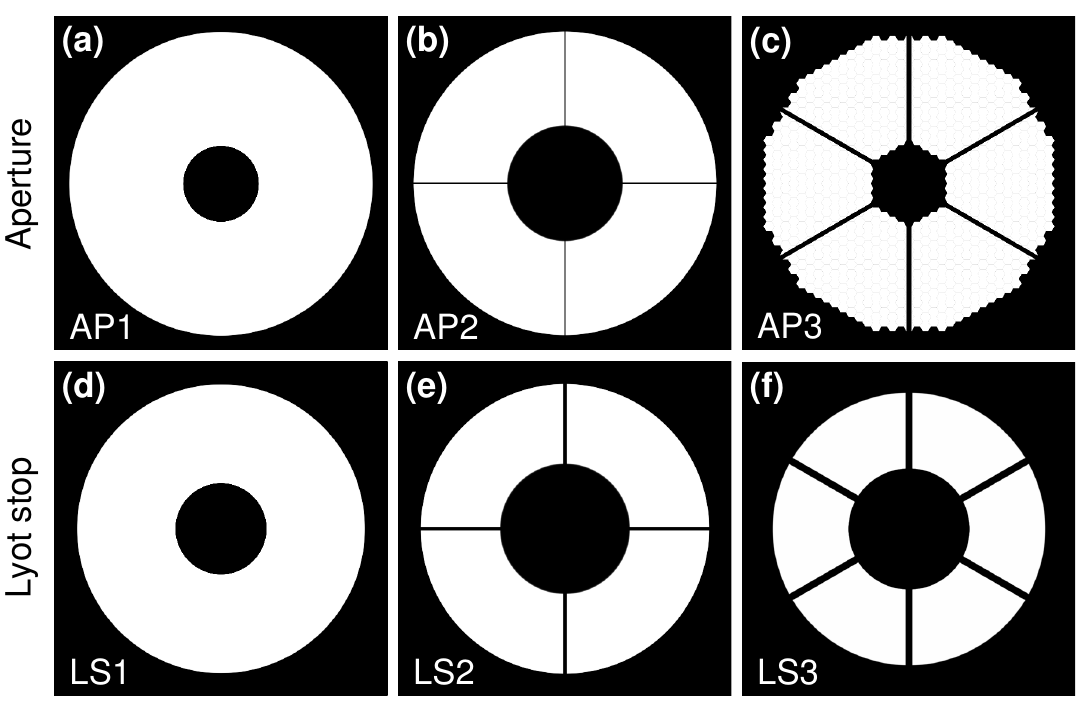}
\end{center}
\caption{ \label{fig:apertures} 
Example telescope apertures and Lyot stops. }
\end{figure} 

\newpage

The optical designs we consider here may be described as a subset of the 4-$f$ optical system shown in Fig.~\ref{fig:generalschematic}. The three relevant planes are the entrance pupil (PP1), focal plane (FP1), and the Lyot plane (PP2). The motivation of this work is to exploit each of these planes by varying the amplitude and phase of the incident light in order to precisely remove the on-axis starlight and improve sensitivity to dim off-axis companions. 

Each optic in the system plays a unique role. The mask at PP1 is used to alter the PSF at FP1. For example, pupil-plane coronagraphs have only one mask at PP1 that forms a dark hole in the PSF. In the case of a focal plane coronagraph, on the other hand, a mask at FP1 blocks the on-axis light or diffracts it outside of the LS at PP2. The focal plane mask and LS are designed to reject the light from the on-axis source, while allowing light from off-axis sources to propagate to the final (or ``science'') image plane. 

Here, we present optimized optical elements in PP1, FP1, and/or PP2 for VCs on telescopes with typical aperture obstructions (see Fig.~\ref{fig:apertures}). The goal is to improve contrast performance without considerable loss in off-axis planet light or image quality. The outline of this manuscript is as follows: section \ref{sec:VC} introduces the VC, section \ref{sec:FPC} discusses correcting optics for FP1 designed to improve starlight rejection, section \ref{sec:PPC} presents complex ``apodizers'' for PP1 and PP2, section \ref{sec:concl} provides conclusions and offers future outlook. In the case of PP1, analytical expressions are given for entrance pupil apodizers that assist in diffracting starlight outside of the LS and potential routes to numerically optimized complex masks are proposed. Lyot plane masks (LPMs) for PP2 are also presented, which improve contrast by relocating leaked starlight in FP2 away from a pre-defined discovery region. The main outcome of this work is that optics in all three planes may be jointly optimized to achieve very high performance with complicated pupils. 

\begin{figure}[t!]
\begin{center}
\includegraphics{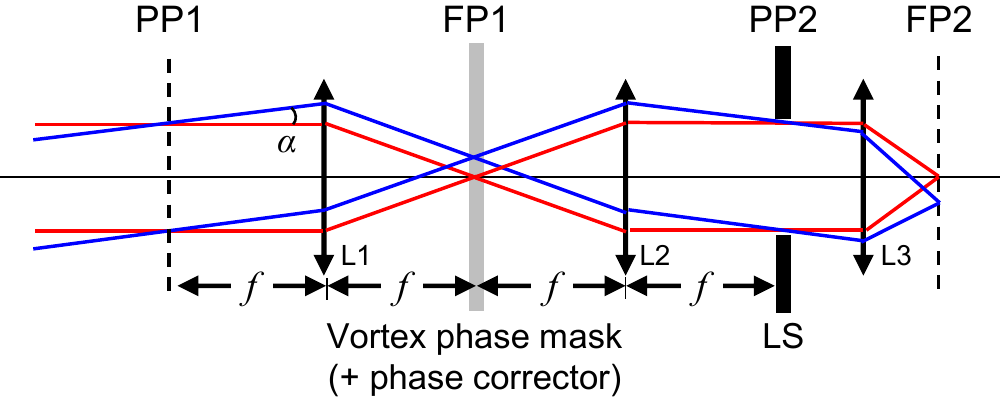}
\end{center}
\caption{ \label{fig:schematic1} 
A vortex coronagraph with focal plane corrector. }
\end{figure} 

\section{THE VORTEX CORONAGRAPH}\label{sec:VC}
In this section, we briefly review the conventional VC where the pupil is circular and has no obstructions. The layout of the VC is illustrated in Fig. \ref{fig:schematic1}. A phase mask is placed in the focal plane with transmission $t(\phi)=\exp(il\phi),$ where $l$ is a nonzero even integer known as ``topological charge" and $\phi$ is the azimuthal angle in FP1. The field immediately after the vortex phase mask owing to an on-axis point source and entrance pupil function $P\left(r,\theta \right) = \mathrm{circ}\left(r/a\right)$ may be written
\begin{equation}
F\left(\rho,\phi \right)=\frac{k a^2}{f}\frac{J_1\left( k a \rho/f\right)}{k a \rho/f} e^{il\phi},
\label{eq:PSF}
\end{equation}
where $\left( r, \theta \right)$ and $\left( \rho, \phi \right)$ are respectively the pupil and focal plane polar coordinates, $a$ is the pupil radius, $k = 2\pi/\lambda,$ $\lambda$ is the wavelength, $f$ is the focal length, and $J_1$ is the Bessel function of the first kind. The field at the output pupil (PP2) is given by the Fourier transform of Eq. \ref{eq:PSF}:
\begin{equation}
E\left(r,\theta \right)=e^{il\theta}\frac{ka}{f}\int\limits_0^\infty{J_1\left(k a \rho/f\right)J_l\left(k r \rho/f\right)d\rho}.
\label{eq:pp2integral}
\end{equation}
Eq. \ref{eq:pp2integral} is related to the discontinuous integral of Weber and Schafheitlin \cite{Watson1922}. For $l$ is nonzero and even,
\begin{equation}
E\left(r,\theta \right)=\left\{ \begin{matrix}
   0 & r \le a  \\
   \frac{a}{r}R_{|l|-1}^1\left( \frac{a}{r} \right){e^{il\theta}} & r>a  \\
\end{matrix} \right. ,
\label{eq:circularROFs}
\end{equation}
where $R_n^m$ are the radial part of the Zernike polynomials \cite{Carlotti2009}. In the case of $l =2$, for example, $E\left(r,\theta \right)=(a/r)^2e^{i2\theta}$ for $r>a$. Remarkably, the field is zero-valued within the geometric image of the pupil. The same circular area of destructive interference (i.e. a ``nodal'' area) appears for all even, nonzero values of $l$. An LS with radius less than $a$, is placed in PP2 to block all of the light from a distant on-axis point source. Off-axis sources do not form a nodal area and are partially transmitted through the LS. Thus, the vortex phase element provides ideal suppression of a distant on-axis point source in the case of a circular pupil. However, telescope apertures are often much more complicated, which may cause a significant amount of on-axis starlight leak through the LS \cite{Jenkins2008}. 

We define the relative transmitted energy as
\begin{equation}
T(\alpha)=\frac{1}{T_0}\int_{\mathrm{PP2}}\left|\tilde{E}\left(r,\theta;\alpha \right) \right|^2 L\left(r,\theta \right) dA,
\label{eq:transdef}
\end{equation}
where $\tilde{E}\left(r,\theta;\alpha \right)$ is the field at PP2 owing to a point source displaced from the optical axis by angle $\alpha$ (i.e. $\tilde{E}\left(r,\theta;\alpha=0 \right)=E\left(r,\theta \right)$), $L\left(r,\theta \right)$ in the binary LS transmission function, $dA$ is the differential area in PP2, and $T_0$ is the transmitted energy due to a point source without a focal plane mask in place.

\begin{figure}[t!]
\begin{center}
\includegraphics[trim =0mm 1mm 1.5mm 0mm,clip=true]{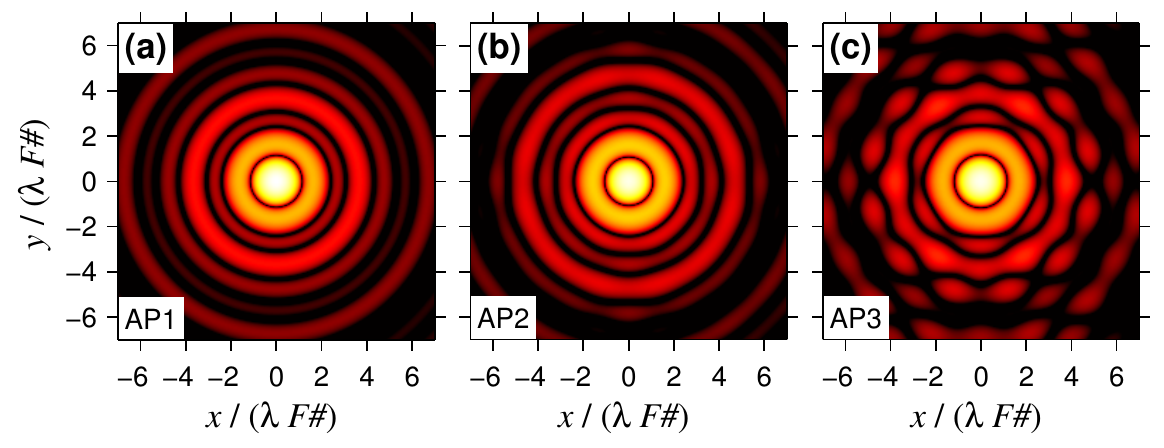}
\includegraphics[trim =53mm -8mm 1mm 8mm,clip=true]{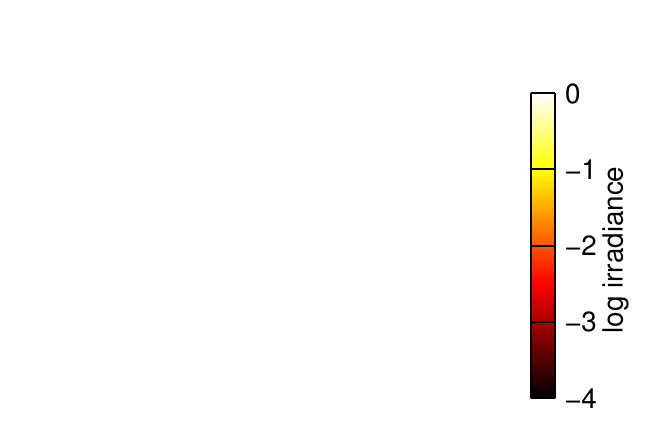}
\end{center}
\caption{ \label{fig:PSFs} 
Monochromatic point spread functions at FP1 in log irradiance for (a) AP1, (b) AP2, and (c) AP3. }
\end{figure}

\section{FOCAL PLANE PHASE CORRECTORS}\label{sec:FPC}

The starlight suppression achieved on telescopes with non-circular or obstructed pupils may be improved by modifying the focal plane phase transmission function. For example, an elliptical aperture requires a skewed vortex phase mask and elliptical LS\cite{Ruane2013}. For arbitrary apertures, we determine the phase function required to form a nodal area in the coherent starlight within the transmitting regions of the LS using a point-by-point iterative phase retrieval algorithm based on Gerchberg-Saxton style error reduction \cite{GerchSax1972,Ruane2015}. 

\subsection{Algorithm}
The field at FP1 $F_j\left(x,y\right)$ is initially taken to be
\begin{equation}
F_1\left(x,y\right)=DFT\{P\left(\xi,\eta\right)\}e^{il\phi},
\end{equation}
where $DFT$ denotes the discrete Fourier transform, $j$ is the iteration number, and $\left( \xi, \eta \right)$ and $\left( x, y \right)$ are discrete Cartesian coordinates in the pupil and focal plane, respectively. The Lyot plane field at each iteration is calculated by 
\begin{equation}
E_j\left(\xi,\eta\right)=DFT^{-1}\{F_j\left(x,y\right)\},
\end{equation}
which is set to zero within the Lyot stop: 
\begin{equation}
\hat{E}_j\left(\xi,\eta\right)=E_j\left(\xi,\eta\right) \left[ 1 - L\left(\xi,\eta \right) \right].
\end{equation}
Then, the new focal plane field is calculated by 
\begin{equation}
G_j\left(x,y\right)=DFT\{\hat{E}_j\left(\xi,\eta\right)\}
\end{equation}
and the returned field magnitude is replaced by the magnitude of the known PSF: 
\begin{equation}
F_{j+1}\left(x,y\right)=|F_1\left(x,y\right)|\;e^{i\mathrm{Arg}\left\{G_j\left(x,y\right)\right\}}. 
\end{equation}
This process is repeated until the starlight energy leaked through the LS in minimized. The updated focal plane correcting mask at each iteration is given by 
\begin{equation}
t_{j}\left(x,y\right)=F_{j}\left(x,y\right)/F_1\left(x,y\right).
\end{equation}

\subsection{Results}
\begin{figure}[t!]
\begin{center}
\includegraphics[width=0.94\linewidth,trim =0mm 1mm 1.5mm 0mm,clip=true]{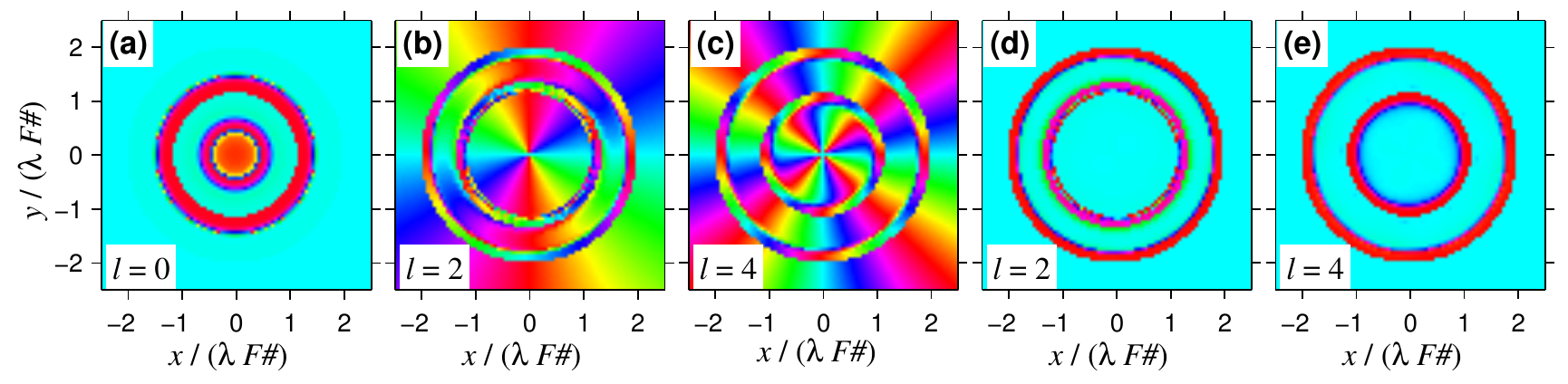}
\includegraphics[trim =53mm -4mm 2mm 9mm,clip=true]{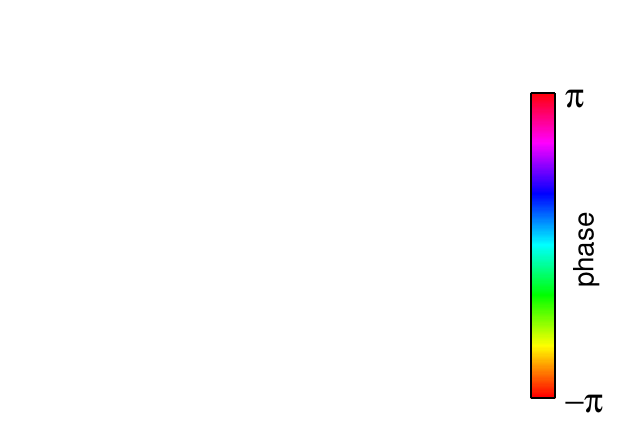}
\end{center}
\caption{ \label{fig:annulus_FPmasks_centeronly} 
Focal plane phase functions optimized for AP1 within a $2~\lambda F\#$ radius using (a) $l=0$, (b) $l=2$, and (c) $l=4$ for the initial condition. (d)-(e)~The corresponding phase correctors for (d) $l=2$ and (e) $l=4$. }
\end{figure}

\begin{figure}[t!]
\begin{center}
\includegraphics[width=0.94\linewidth,trim =0mm 1mm 1.5mm 0mm,clip=true]{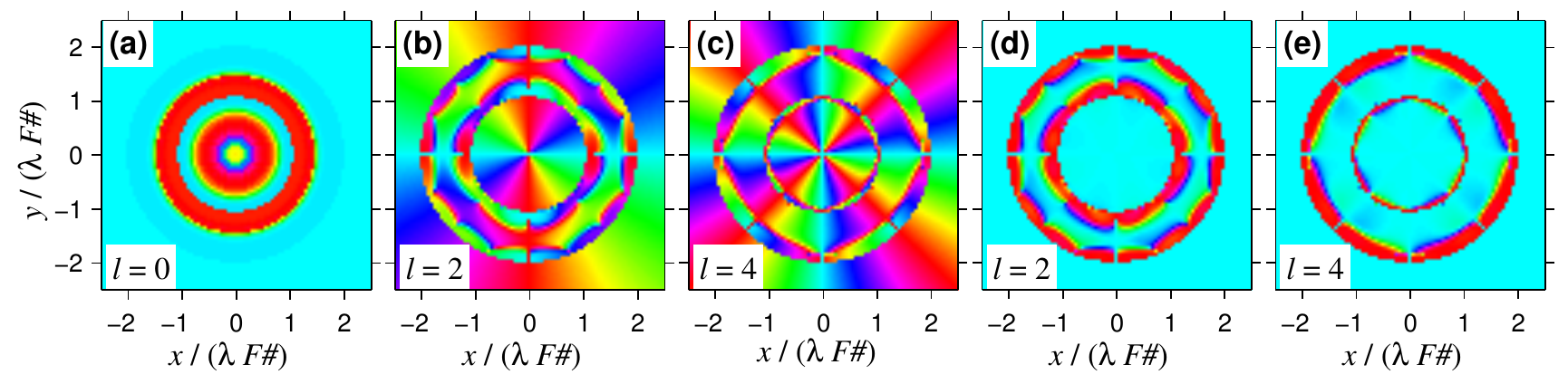}
\includegraphics[trim =53mm -4mm 2mm 9mm,clip=true]{vertphasecolorbar.pdf}
\end{center}
\caption{ \label{fig:palomar_FPmasks_centeronly} 
Same as Fig. \ref{fig:annulus_FPmasks_centeronly}, but for AP2.} 
\end{figure}

\begin{figure}[t!]
\begin{center}
\includegraphics[width=0.94\linewidth,trim =0mm 1mm 1.5mm 0mm,clip=true]{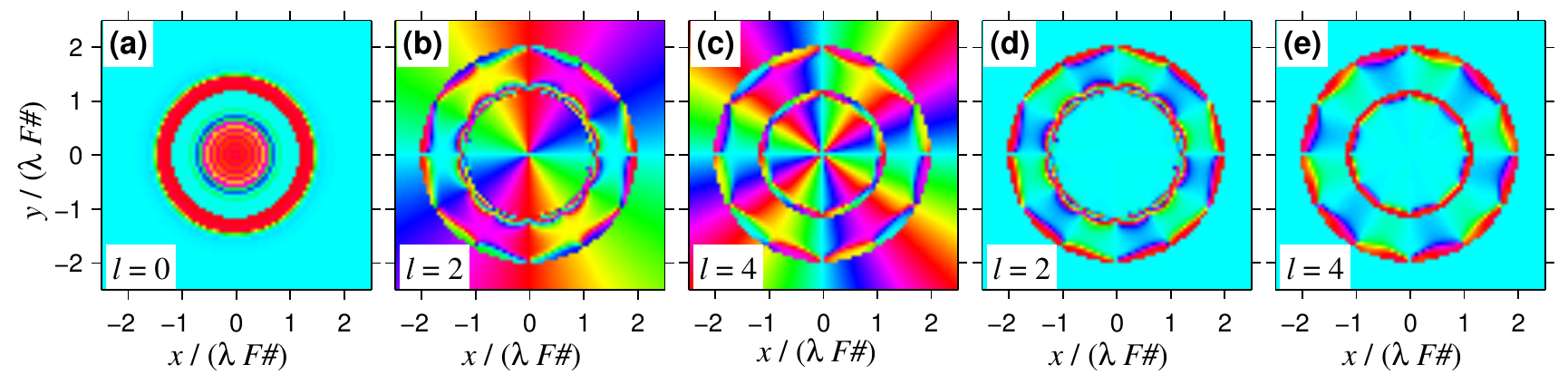}
\includegraphics[trim =53mm -4mm 2mm 9mm,clip=true]{vertphasecolorbar.pdf}
\end{center}
\caption{ \label{fig:ELT_FPmasks_centeronly} 
Same as Figs. \ref{fig:annulus_FPmasks_centeronly} and \ref{fig:palomar_FPmasks_centeronly} , but for AP3.} 
\end{figure}

Figure \ref{fig:PSFs} shows the PSFs associated with the pupils in Fig. \ref{fig:apertures} in terms of normalized irradiance: $\left|F_1\left(x,y\right)\right|^2$. We wish to diffract most of the light in the on-axis PSF outside of the LS (see Fig. \ref{fig:apertures}(d)-(f)), while preserving light from off-axis sources at angular separations as small as $\alpha \approx 2-3 \lambda/D$, where $D$ is the outer diameter of the entrance aperture. With this in mind, we chose to optimize a phase corrector that only modifies the focal plane phase within a $2~\lambda F\#$ radius of the center, where $F\#=f/D$. This constraint limits unwanted suppression of light from sources at $\alpha \gtrsim 2~\lambda/D$. The calculated focal plane corrections are shown in Figs. \ref{fig:annulus_FPmasks_centeronly}--\ref{fig:ELT_FPmasks_centeronly} for AP1, AP2, and AP3, respectively. Throughout this work, we perform the DFTs using the Fast Fourier Transform algorithm with a $16,384\times16,384$ computational grid of samples and with $\sim\!\!16$ samples per $\lambda F\#$ in FP1. In each case, we find the optimized phase corrector for an initial phase mask with $l=0,~2,~\mathrm{and}~4$. For the $l=0$ case, we force circular symmetry every 10 iterations until about 50 iterations to encourage convergence to a circularly symmetric mask. This constraint was also applied for all of the correctors calculated for AP1. The more complicated apertures (AP2 and AP3) naturally lead to more intricate solutions. The algorithm is stopped after 500 iterations.

\begin{figure}[t!]
\begin{center}
\includegraphics{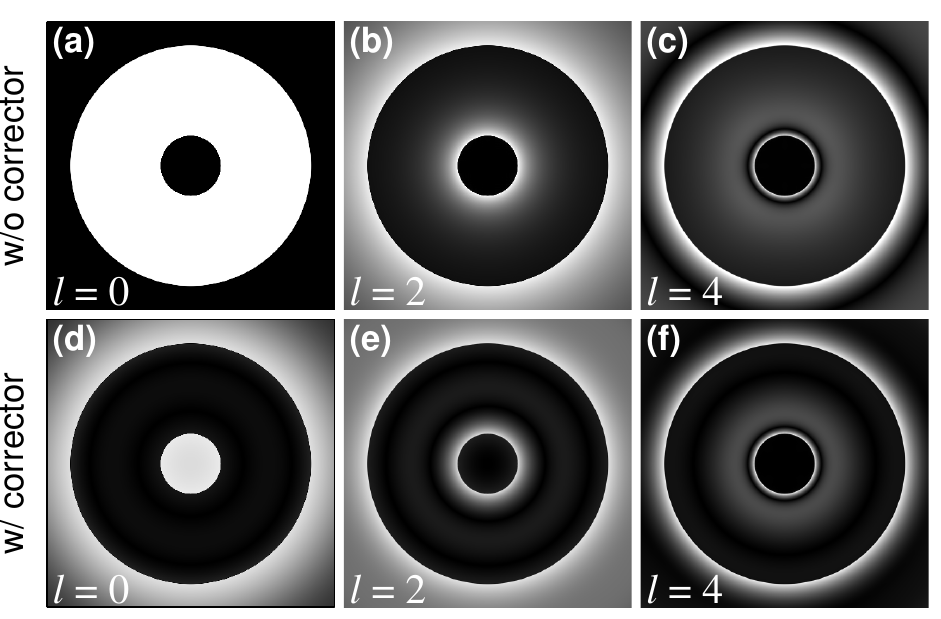}
\includegraphics[trim =53mm -4mm 2mm 11mm,clip=true]{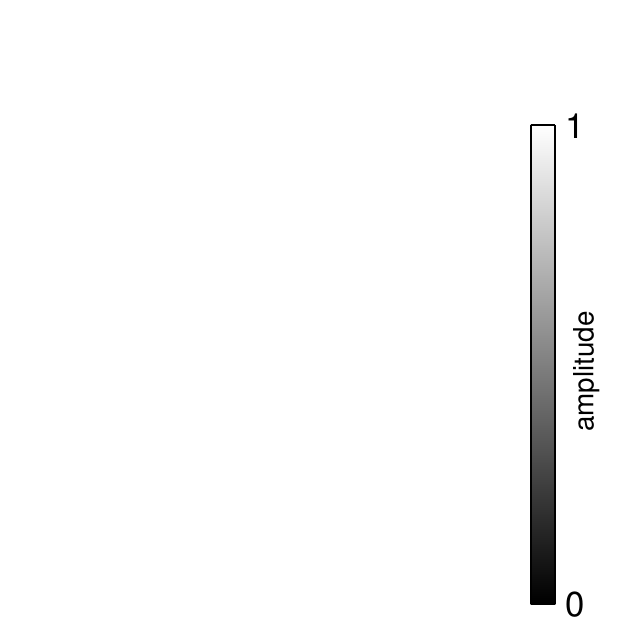}
\end{center}
\caption{ \label{fig:ROFs_Annulus} 
Lyot plane field amplitude for AP1 (a)-(c) without and (d)-(f) with a focal plane phase corrector. }
\end{figure} 

\begin{figure}[t!]
\begin{center}
\includegraphics{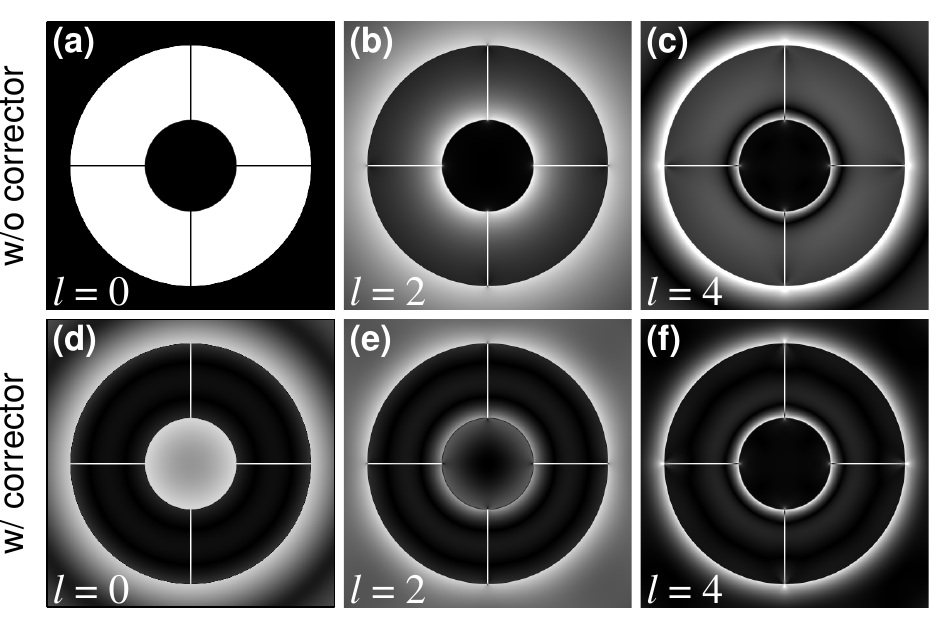}
\includegraphics[trim =53mm -4mm 2mm 11mm,clip=true]{vertabscolorbar2.pdf}
\end{center}
\caption{ \label{fig:ROFs_Palomar} 
Same as Fig. \ref{fig:ROFs_Annulus}, but for AP2. }
\end{figure} 

\begin{figure}[t!]
\begin{center}
\includegraphics{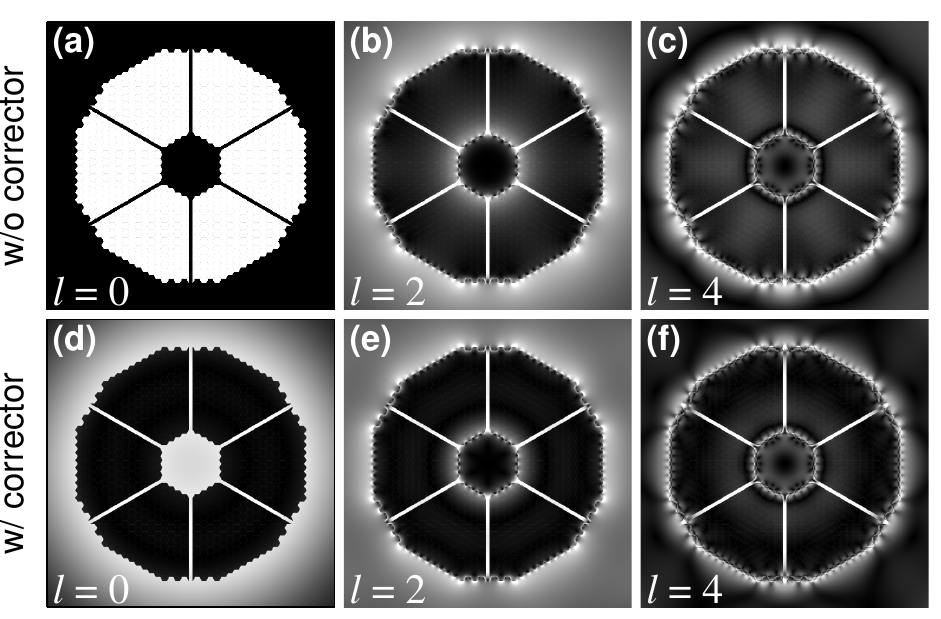}
\includegraphics[trim =53mm -4mm 2mm 11mm,clip=true]{vertabscolorbar2.pdf}
\end{center}
\caption{ \label{fig:ROFs_ELT} 
Same as Fig. \ref{fig:ROFs_Annulus} and \ref{fig:ROFs_Palomar}, but for AP3. }
\end{figure} 

The algorithm typically finds a local optimum close to the initial condition, which is a beneficial property for our application where contrast improvement is desired with minimal phase shifts and phase mask complexity. The correctors generally take the form of concentric rings with radial phase steps of approximately $\pi$. Azimuthal variations in the phase corrector are present in cases where it is necessary to mitigate the effect of the spiders and/or non-circular pupil as in AP2 and AP3. We also note that non-circularly symmetric solutions may be obtained for circularly symmetric apertures. 

The magnitude of the Lyot plane field owing to a on-axis point source for each design is shown in Figs. \ref{fig:ROFs_Annulus}--\ref{fig:ROFs_ELT}. In all cases, most of the light is diffracted outside of the geometric image of the entrance pupil (equivalent to Figs. \ref{fig:ROFs_Annulus}--\ref{fig:ROFs_ELT}(a)). This provides a spatial separation of the light from on-axis and off-axis sources allowing the LS to block only the on-axis source.

\begin{figure}[t!]
\begin{center}
\includegraphics[width=\linewidth,trim =0mm 0mm 0mm 0mm,clip=true]{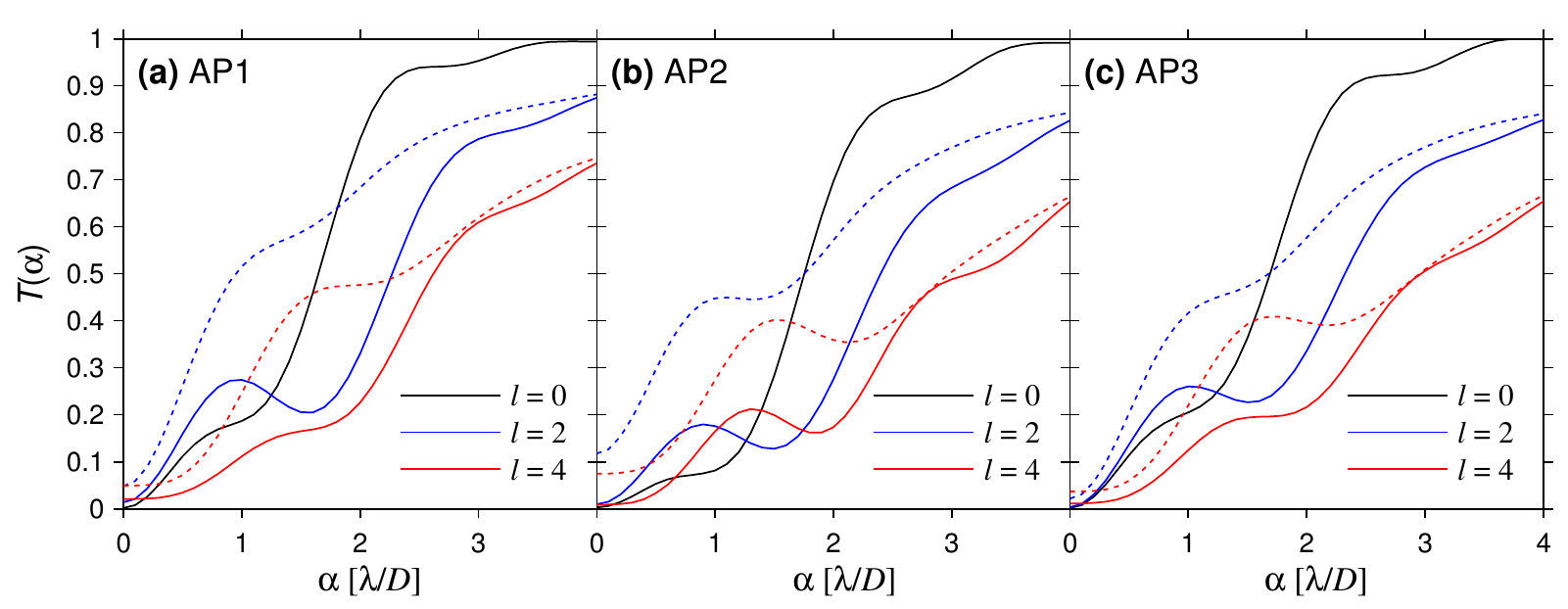}
\end{center}
\caption{ \label{fig:offaxisTransPower} 
Transmitted energy for a distant point source at angular displacement $\alpha$ with (solid lines) and without (dotted lines) the focal plane correctors for (a) AP1, (b) AP2, and (c) AP3. } 
\end{figure} 

\begin{table}[b]
\caption{The on-axis starlight suppression $T\left(0\right)$ and inner working angle (IWA) in units of $\lambda/D$.} 
\label{tab:OffAxisPerformance}
\begin{center}       
\begin{tabular}{ c|c|c|c|c|c|c|c|c|c|c|c|c|} 
\cline{2-13}
\rule[-1ex]{0pt}{3.5ex}
  & \multicolumn{4}{|c|}{AP1} & \multicolumn{4}{|c|}{AP2} & \multicolumn{4}{|c|}{AP3}  \\
\cline{2-13}
\rule[-1ex]{0pt}{3.5ex}  
 & \multicolumn{2}{|c|}{w/o corrector} & \multicolumn{2}{|c|}{w/ corrector} & \multicolumn{2}{|c|}{w/o corrector} & \multicolumn{2}{|c|}{w/ corrector} & \multicolumn{2}{|c|}{w/o corrector} & \multicolumn{2}{|c|}{w/ corrector} \\
\cline{2-13} 
\rule[-1ex]{0pt}{3.5ex} 
 & $T\left(0\right)$ & IWA & $T\left(0\right)$ & IWA & $T\left(0\right)$ & IWA& $T\left(0\right)$ & IWA & $T\left(0\right)$ & IWA & $T\left(0\right)$ & IWA\\
\hline
\multicolumn{1}{|c|}{$l=0$} & 1.0 & N/A & $0.0015$ & 1.65 & 1.0 & N/A & 0.0028 & 1.76 & 1.0 & N/A & 0.0015 & 1.67   \\
\hline
\multicolumn{1}{|c|}{$l=2$} & 0.0481 & 0.95 & 0.0136 & 2.27 & 0.1172 & 1.75 & 0.0093 & 2.40 & 0.0214 & 1.68 & 0.0034 & 2.33   \\
\hline
\multicolumn{1}{|c|}{$l=4$} & 0.0493 & 2.35 & 0.0203 & 2.62 & 0.0742 & 2.98 & 0.0084 & 3.15 & 0.0363 & 2.95 & 0.0112 & 2.97   \\
\hline 
\end{tabular}
\end{center}

\end{table} 

\subsection{Performance}

We compare the performance of the solutions presented above in terms of starlight suppression, off-axis transmission, and broadband performance. For each initial condition, the algorithm arrives at a phase corrector that provides a different starlight suppression level $T\left(0\right)$, which can be seen in Fig. \ref{fig:offaxisTransPower}. The values of $T\left(0\right)$ are reported in Table 1. Generally, the $l=0$ initial condition achieves better suppression of the on-axis source. As explained below, the advantage of using $l=2~\mathrm{and}~4$ is reduced sensitivity to low-order aberrations, vibrations, partial resolution of the star, and chromatic effects. We also find that based on the distribution of leaked light in Figs. \ref{fig:ROFs_Annulus}--\ref{fig:ROFs_ELT}, a larger LS inner radius is preferred for $l=2~\mathrm{and}~4$ where a ring of light around the secondary mirror is present. This explains the enhanced performance of the $l=2$ solution for AP3. 

\subsubsection{Off-axis transmission}
One side-effect of the phase corrector is that it reduces light from off-axis sources near to the parent star. Thus, in addition to reducing the light from the on-axis star, we wish to maximize the signal detected from sources of interest. Figure \ref{fig:offaxisTransPower} shows the transmitted energy for off-axis sources $T\left(\alpha\right)$ (defined in Eq. \ref{eq:transdef}). The inner working angle of the coronagraph is defined as the angle at which half of the planet signal is transmitted: $T\left(\mathrm{\alpha=IWA}\right)=0.5$. The IWAs achieved in each case are reported in Table 1. 

Introducing phase corrections only within $2~\lambda F\#$ allows for an $IWA\approx2-3~\lambda/D$ in most cases. Using a larger phase corrector leads to a smaller value for $T\left(0\right)$, but may increase the IWA considerably. We note that it is possible to design phase corrections in an arbitrary region of the focal plane, including asymmetric shapes where the IWA varies azimuthally. 

One attractive property of using an $l=4$ solution is that it is less sensitive to very small displacements, which can be seen in in Fig. \ref{fig:offaxisTransPower}. This is a benefit when vibration and/or other small tip-tilt errors are present. Not only is the $l=4$ case more robust to aberrations (tip-tilt and defocus) as compared to $l=0~\mathrm{and}~2$, it is also less sensitive to the finite angular size of the star \cite{Mawet2010b,Delacroix2014}. 

\begin{figure}[t!]
\begin{center}
\includegraphics[width=\linewidth,trim =0mm 0mm 0mm 0mm,clip=true]{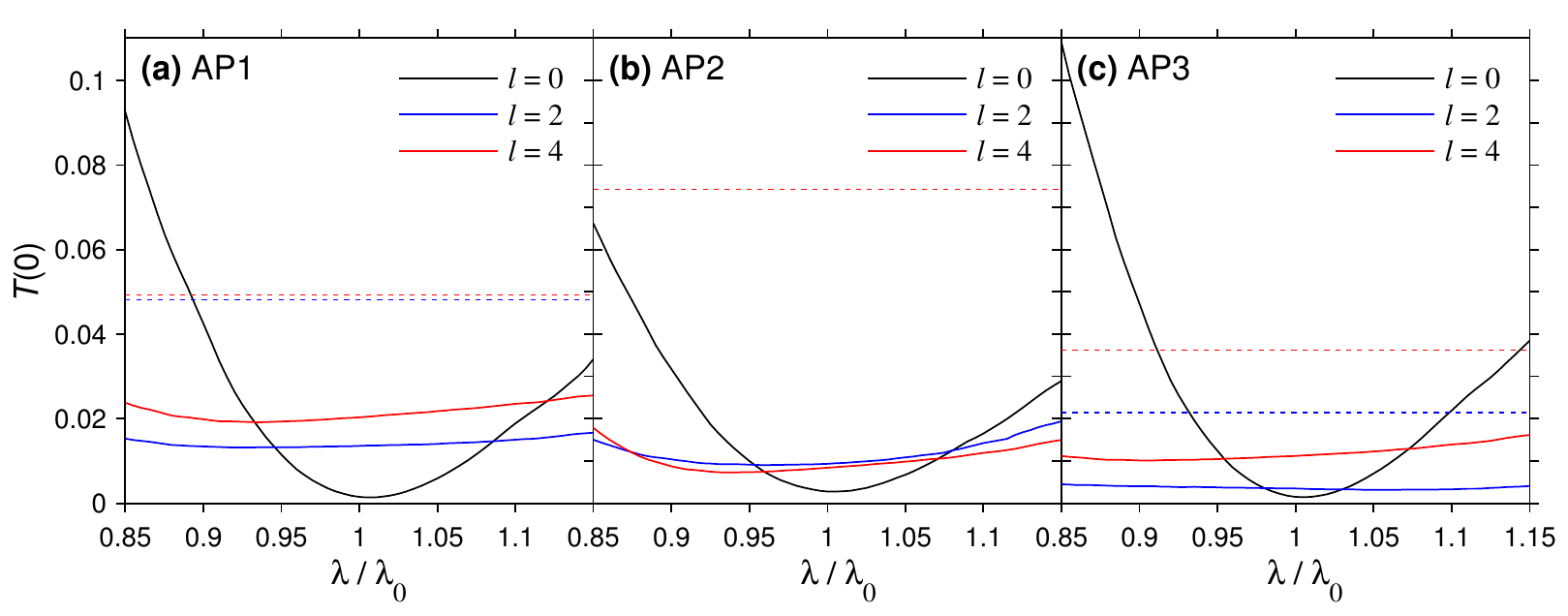}
\end{center}
\caption{ \label{fig:chromaticity} 
Transmitted energy for an on-axis point source as a function of wavelength $\lambda$ with (solid lines) and without (dotted lines) the focal plane correctors for (a) AP1, (b) AP2, and (c) AP3. $\lambda_0$ is the design wavelength. } 
\end{figure} 

\subsubsection{Wavelength dependence}
A potential benefit of using vortex phase masks is that they are intrinsically achromatic. That is, assuming perfect optics, the expected field at the LS (see Figs. \ref{fig:ROFs_Annulus}--\ref{fig:ROFs_ELT}) does not vary with wavelength $\lambda$. The phase correctors, on the other hand, have radial variations and are designed for a specific wavelength $\lambda_0$. Figure \ref{fig:chromaticity} shows the transmitted energy through the LS $T\left(\alpha\right)$ as a function of $\lambda$. As previously noted, the $l=0$ solutions yield the best starlight suppression at the design wavelength. However, the $l=2~\mathrm{and}~4$ solutions are far less sensitive to wavelength and may provide better broadband starlight suppression, especially in the case of more complicated pupils (AP2 and AP3). 

Achromatic vortex phase masks have been manufactured using subwavelength gratings \cite{Bomzon2001,Mawet2005}, liquid crystals \cite{Marrucci2006,Mawet2009}, and photonic crystals \cite{Murakami2012,Murakami2013}. One example is the annular groove phase mask (AGPM), which is an $l=2$ vortex phase masks that operates in the mid-infrared \cite{Mawet2005,Delacroix2013} (also see Absil et al., these proceedings). AGPMs provide achromatic phase-only transmission via form birefringence. Similar subwavelength gratings to the AGPM can be fabricated to produce a higher charge vortex or potentially even more complicated phase patterns \cite{Delacroix2014}. The liquid crystal and photonic crystal elements may also be used to fabricate high-fidelity vortex phase masks (and other phase patterns) that operate in the visible or infrared wavelength regime. Future work will incorporate methods to reduce the wavelength dependence of the phase corrector.

\subsection{Complex (or ``hybrid") focal plane correctors}

\begin{figure}[t!]
\begin{center}
\includegraphics[width=0.52\linewidth,trim =0mm 1mm 1.5mm 0mm,clip=true]{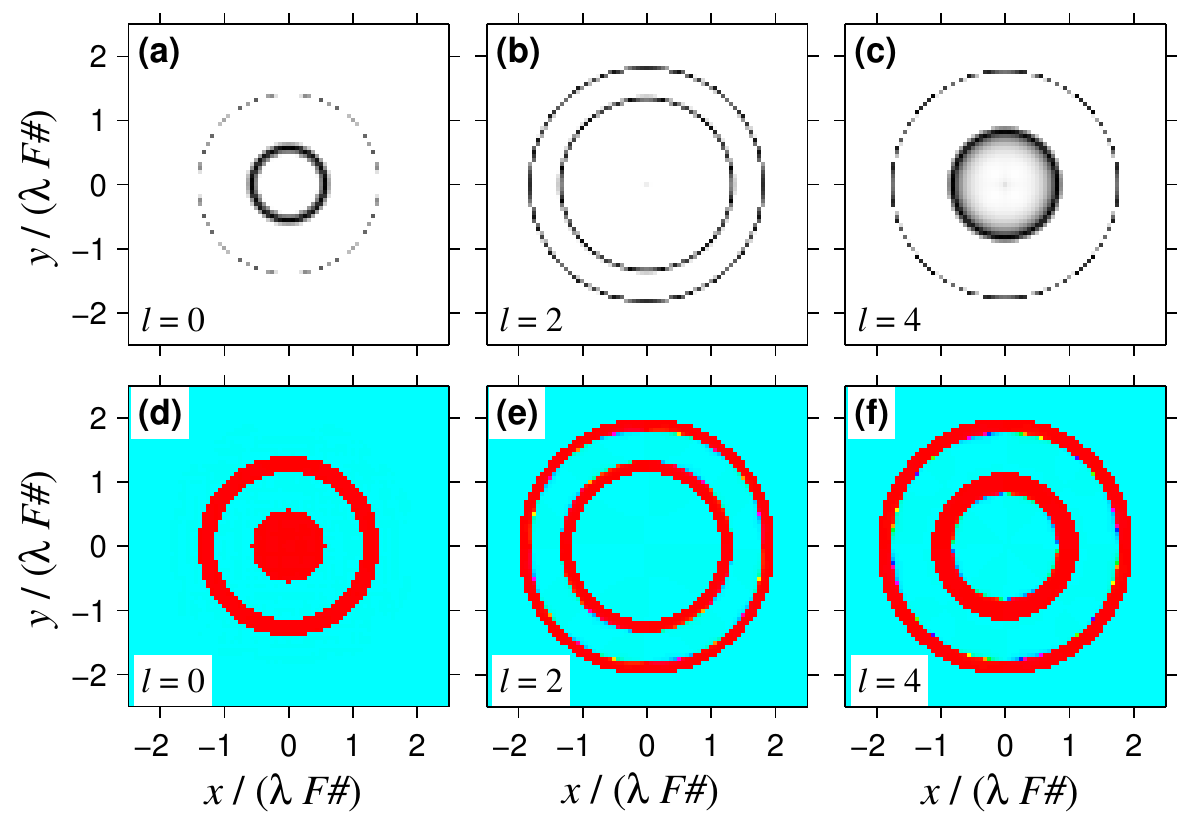}
\includegraphics[trim =50mm -4mm 1mm 2mm,clip=true]{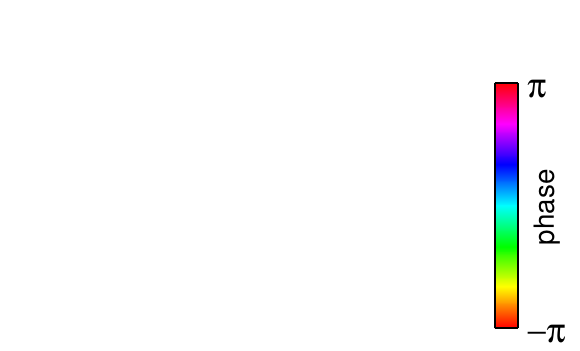}
\end{center}
\caption{ \label{fig:annulus_FPmasks_complex} 
(a)-(c) Amplitude and (d)-(f) phase of complex focal plane correctors optimized for AP1 within a $2~\lambda F\#$ radius using (a),(d) $l=0$, (b),(e) $l=2$, and (c),(f) $l=4$ for the initial condition. }
\end{figure} 

\begin{figure}[t!]
\begin{center}
\includegraphics[width=0.52\linewidth,trim =0mm 1mm 1.5mm 0mm,clip=true]{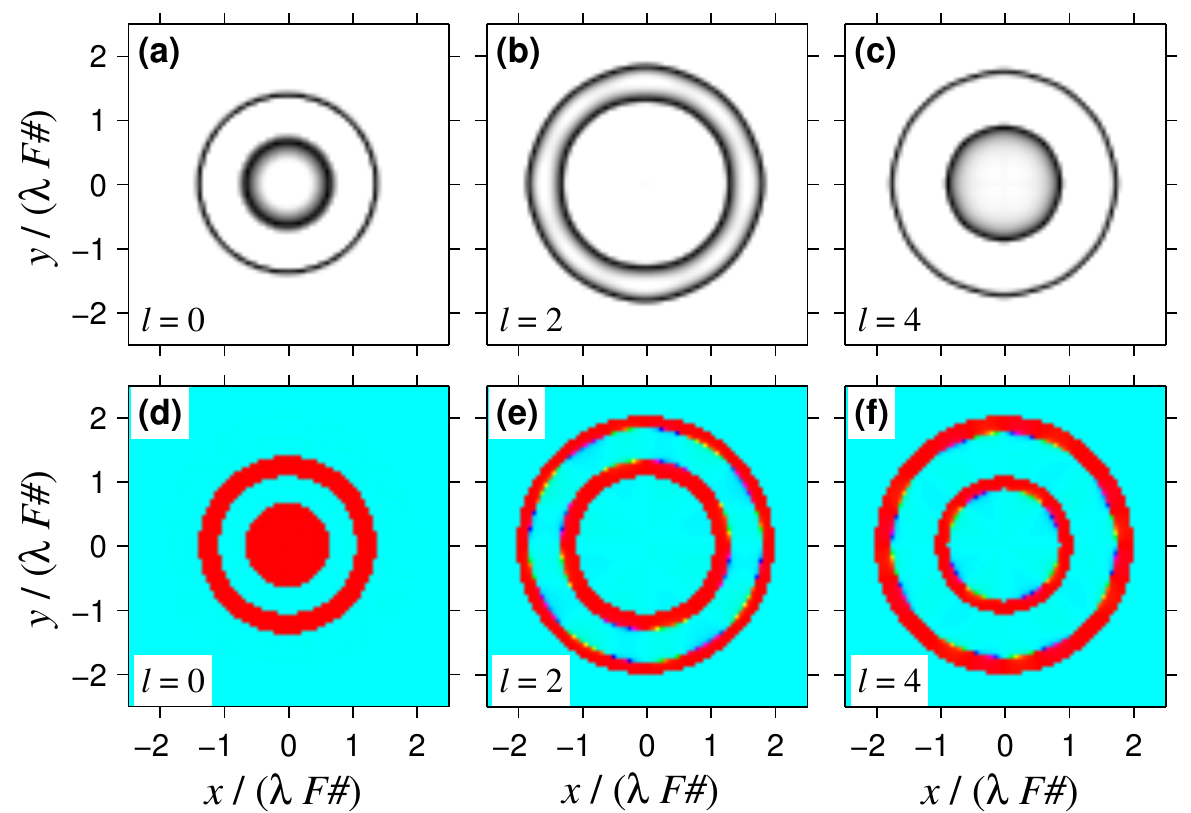}
\includegraphics[trim =50mm -4mm 1mm 2mm,clip=true]{vertphasecolorbar3.pdf}
\end{center}
\caption{ \label{fig:palomar_FPmasks_complex} 
Same as Fig. \ref{fig:annulus_FPmasks_complex}, but for AP2.} 
\end{figure} 

\begin{figure}[t!]
\begin{center}
\includegraphics[width=0.52\linewidth,trim =0mm 1mm 1.5mm 0mm,clip=true]{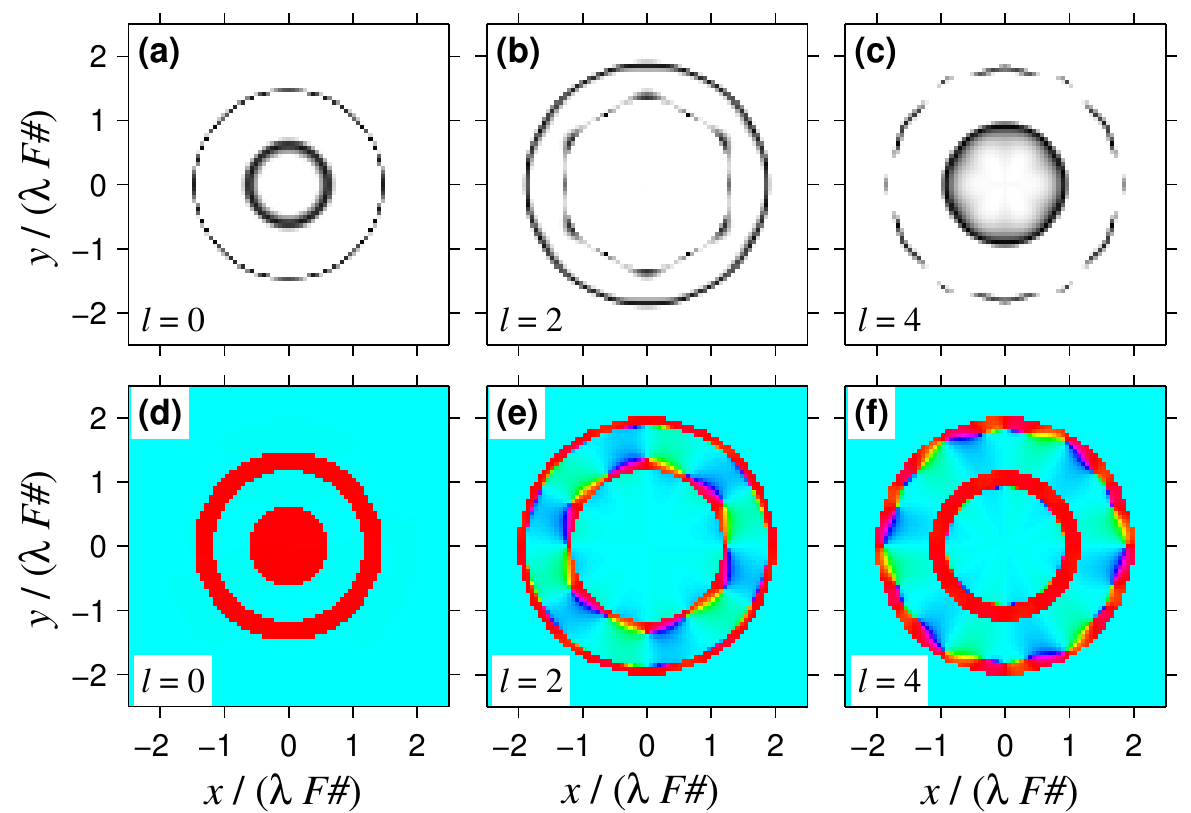}
\includegraphics[trim =50mm -4mm 1mm 2mm,clip=true]{vertphasecolorbar3.pdf}
\end{center}
\caption{ \label{fig:ELT_FPmasks_complex} 
Same as Figs. \ref{fig:annulus_FPmasks_complex} and \ref{fig:palomar_FPmasks_complex} , but for AP3.} 
\end{figure} 

\begin{figure}[t!]
\begin{center}
\includegraphics[width=\linewidth,trim =0mm 0mm 0mm 0mm,clip=true]{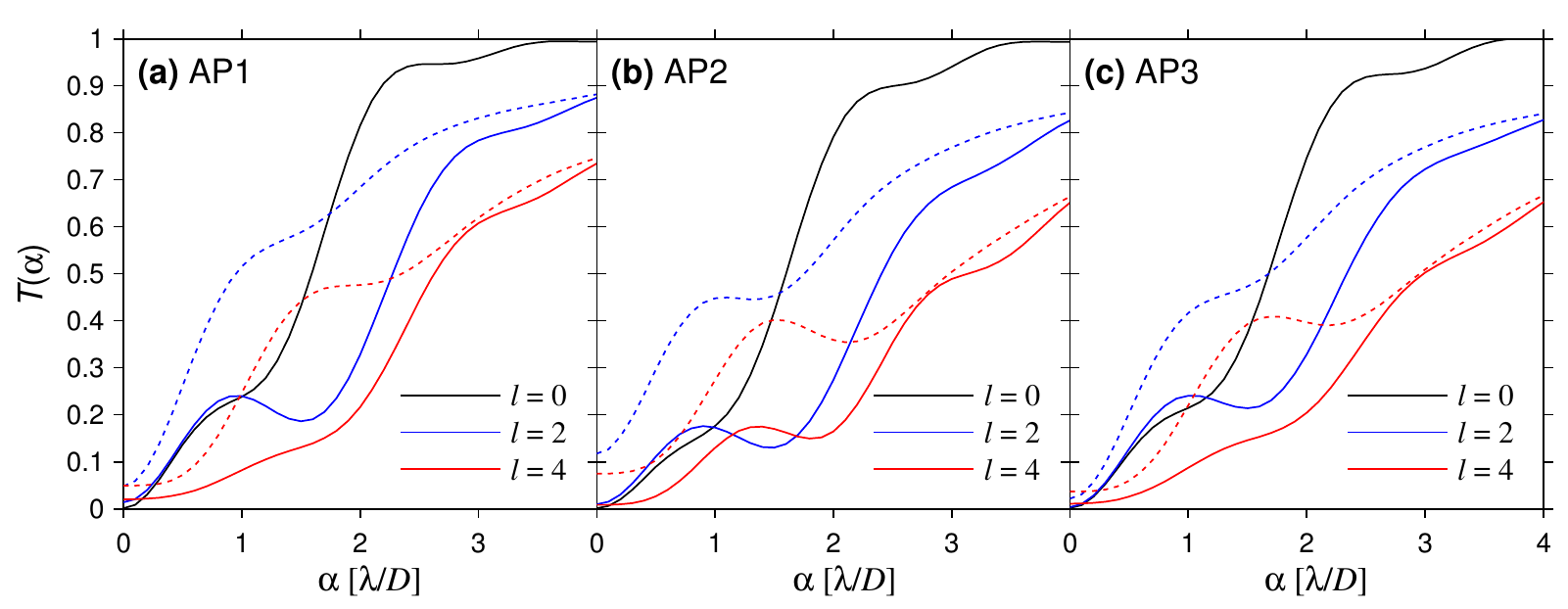}
\end{center}
\caption{ \label{fig:offaxisTransPower_complex} 
Transmitted energy for a distant point source at angular displacement $\alpha$ with (solid lines) and without (dotted lines) the complex focal plane correctors for (a) AP1, (b) AP2, and (c) AP3. } 
\end{figure}

\begin{figure}[t!]
\begin{center}
\includegraphics[width=\linewidth,trim =0mm 0mm 0mm 0mm,clip=true]{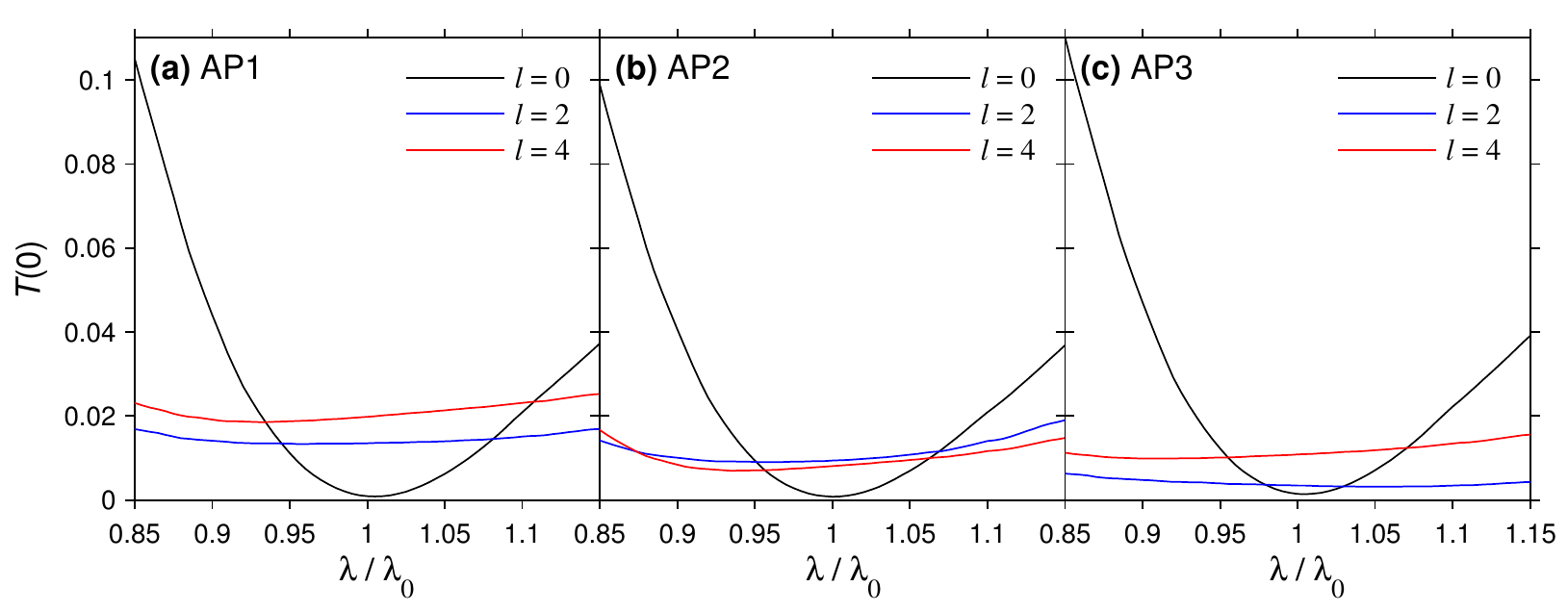}
\end{center}
\caption{ \label{fig:chromaticity_complex} 
Transmitted energy for an on-axis point source as a function of $\lambda$ with complex focal plane correctors for (a) AP1, (b) AP2, and (c) AP3. } 
\end{figure}

Rather than calculating phase-only corrections, it is also possible to use a similar algorithm to optimize both the amplitude and phase. The complex solutions are shown in Figs. \ref{fig:annulus_FPmasks_complex}--\ref{fig:ELT_FPmasks_complex}. The corrections appear as concentric opaque rings in addition to radial phase steps. The phase functions appear slightly less complicated for the complex masks as compared to the phase-only solutions. Depending on the fabrication methods used, it may be more feasible to apply a semi-transparent amplitude function than high frequency phase variations. The off-axis transmission and wavelength dependence are shown in Figs. \ref{fig:offaxisTransPower_complex} and \ref{fig:chromaticity_complex}, respectively. The performance is quite similar to the phase-only case (Figs. \ref{fig:offaxisTransPower}--\ref{fig:chromaticity}), but with slightly reduced values for $T\left(0\right)$. 

\section{OPTIMIZED PUPIL PLANE ELEMENTS}\label{sec:PPC}
An alternate way to achieve improved starlight suppression and contrast performance is to introduce optimized optical elements in the pupil planes of the coronagraph. Here, we discuss using masks located in the entrance pupil and Lyot plane of a VC (as depicted in Fig. \ref{fig:generalschematic}).

\subsection{Zernike amplitude pupil apodizers}

\begin{figure}[t!]
\centering
\includegraphics{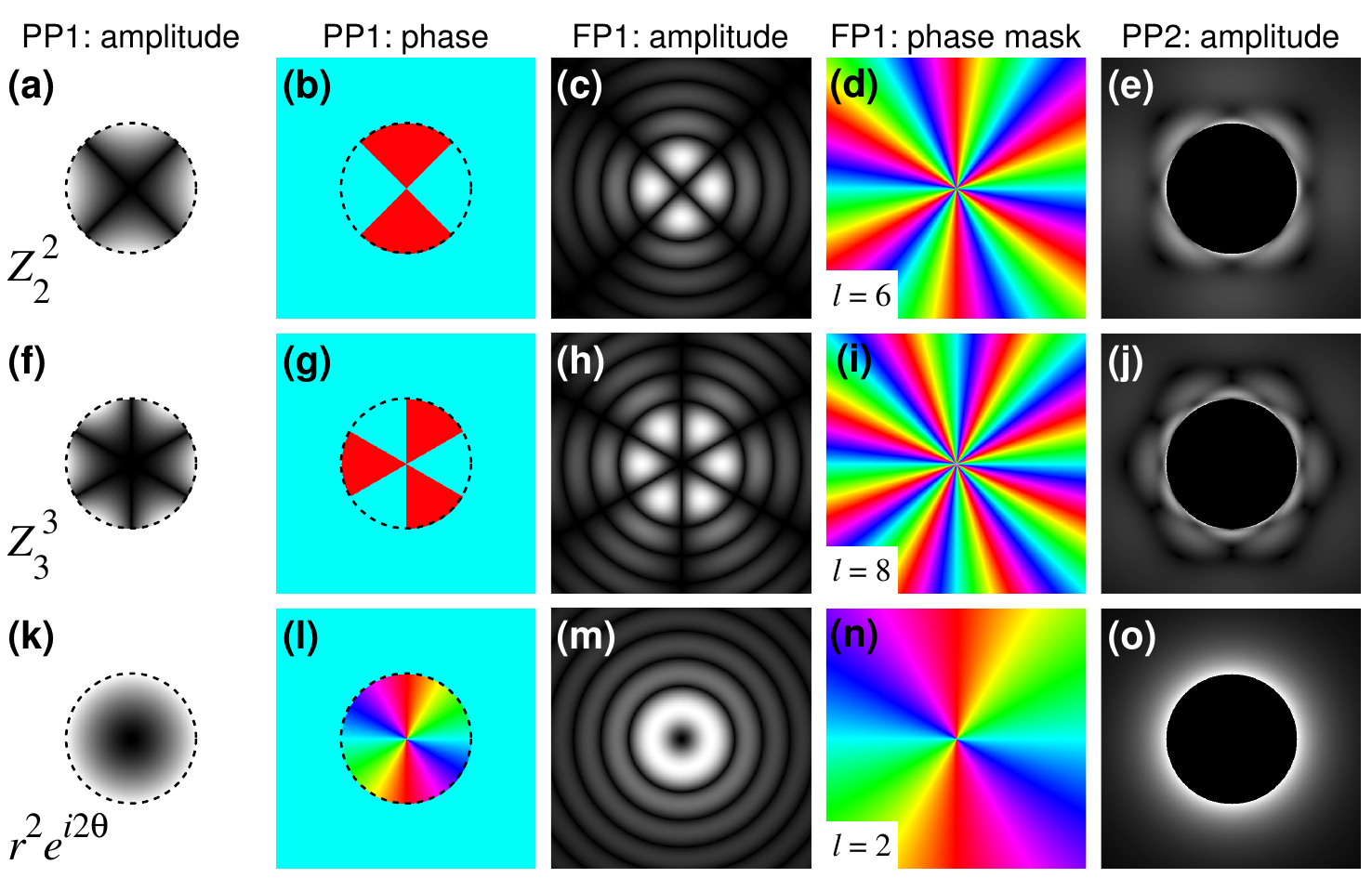}\hfill \\ 
    \begin{subfigure}{0.4\linewidth}
      \hspace{18mm}\includegraphics[scale =0.8,trim = 0 0 0 33mm,clip=true]{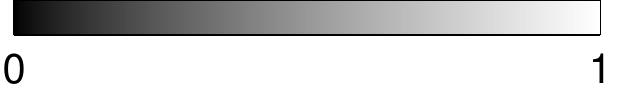}
    \end{subfigure}
    \hspace{10mm}
    \begin{subfigure}{0.4\linewidth}
      \includegraphics[scale =0.8,trim = 0 0 0 31.5mm,clip=true]{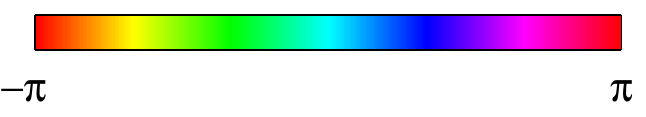}
    \end{subfigure}
\caption{ \label{fig:Zernike} 
Example Zernike amplitude apodizers. (a)-(b)~The pupil amplitude $|P(r,\theta)|$ and phase $\mathrm{Arg}\{P(r,\theta)\}$ in PP1, respectively. (c)~The corresponding point spread function magnitude in FP1 $|F(\rho,\phi)|$. (d)~The vortex phase element in FP1. (e)~The field magnitude just before the LS in PP2 $|E\left(r,\theta \right)|$. The example pupil functions shown are (a)-(e)~$Z_2^2$, (f)-(j)~$Z_3^3$, and (k)-(o)~$Z_2^2+iZ_2^{-2}$. $|F(\rho,\phi)|$ is shown over a $10\times10~\lambda F\#$ square.}
\end{figure} 

In this section, we present an analytical basis for complex apodizers for the VC. The pupil amplitude functions are described by real-valued Zernike polynomials: 
\begin{equation}
P\left(r,\theta\right) = Z_n^m\left(r/a,\theta\right),\;\;\;\;\;r\le a.
\end{equation}
We show that under certain conditions ideal contrast is achieved (see Fig. \ref{fig:Zernike}). For $m\ge0$ (i.e. the even Zernike polynomials),
\begin{equation}
P\left(r,\theta\right) = R_n^m\left(r/a\right) \cos\left(m\theta\right),\;\;\;\;\;r\le a,
\label{eq:Zpupil}
\end{equation}
where $R_n^m\left(r/a\right)$ are the radial polynomials used in Eq. \ref{eq:circularROFs}. The field transmitted through a vortex phase element of charge $l$ in FP1, owing to an on-axis point source, is given by the product of $\exp\left(il\phi\right)$ and the Fourier transform of Eq. \ref{eq:Zpupil}:
\begin{equation}
F\left(\rho,\phi \right)=\frac{k a^2}{f}\frac{J_{n+1}\left( k a \rho/f\right)}{k a \rho/f} \cos\left(m\phi\right) e^{il\phi}.
\label{eq:ZPSF}
\end{equation}
The field in PP2 (just before the LS) is given by the Fourier transform of Eq. \ref{eq:PSF}:
\begin{equation}
E\left(r,\theta \right)=\frac{k a}{2 f}e^{il\theta}\left[(-1)^m e^{im\theta}W_{n+1}^{l+m}(r)+e^{-im\theta}W_{n+1}^{l-m}(r)\right],~m\ge0
\label{eq:Wevens}
\end{equation}
where 
\begin{equation}
W_p^q(r)=\int\limits_{0}^{\infty }{J_p\left({k a \rho}/{f}\right)J_q\left( {k r \rho}/{f}\right)d\rho}.
\label{eq:Wpq}
\end{equation}
Similarly, for $m<0$ (i.e. the odd Zernike polynomials)
\begin{equation}
E\left(r,\theta \right)=\frac{k a}{i2 f}e^{il\theta}\left[(-1)^m e^{im\theta}W_{n+1}^{l+m}(r)-e^{-im\theta}W_{n+1}^{l-m}(r)\right], ~m<0.
\label{eq:Wodds}
\end{equation}
For an on-axis point source, a nodal area appears at PP2 if $|l|>n+|m|$ and $l$ is even valued. Analytical solutions for $E\left(r,\theta \right)$ that contain a nodal area in the on-axis starlight may be written
\begin{equation}
E\left(r,\theta \right)=\left\{ \begin{matrix}
   0 & r \le a  \\
   g^{(l)}_{n,m}\left(r,\theta \right) & r>a  \\
\end{matrix} \right. .
\label{eqn:outsideg}
\end{equation}
Figure \ref{fig:Zernike} shows three example field patterns at PP2 for three relevant Zernike amplitude apodizers. The analytical expressions of $g^{(l)}_{n,m}\left(r,\theta \right)$ for the examples shown are

\begin{equation}
\begin{split}
g^{(6)}_{2,2}\left(r,\theta \right)&=\left[\frac{21}{2}\left(\frac{a}{r}\right)^8 -15\left(\frac{a}{r}\right)^6 + 5\left(\frac{a}{r}\right)^4 \right]e^{i8\theta}+\frac{1}{2}\left(\frac{a}{r}\right)^4 e^{i4\theta},\\
g^{(8)}_{3,3}\left(r,\theta \right)&=\left[60\left(\frac{a}{r}\right)^{11}-126\left(\frac{a}{r}\right)^9+84\left(\frac{a}{r}\right)^7-\frac{35}{2}\left(\frac{a}{r}\right)^5\right]{e^{i11\theta}}+\frac{1}{2}\left(\frac{a}{r}\right)^5{e^{i5\theta}}.
\end{split}
\end{equation}
We also present the special case where 
\begin{equation}
\begin{split}
P\left(r,\theta\right) &= Z_m^m\left(r/a,\theta\right) \pm iZ_m^{-m}\left(r/a,\theta\right),\;\;\;\;\;r\le a,\\
 &= \left(r/a\right)^{m}e^{im\theta},\;\;\;\;\;r\le a.
 \end{split}
\end{equation}
For $m\ge0$, it can be shown that for all even values of $l>0$, the Lyot plane field becomes
\begin{equation}
E\left(r,\theta \right)=\left\{ \begin{matrix}
   0 & r \le a  \\
   \left(\frac{a}{r}\right)^{l+m}e^{i\left(l+m\right)\theta} & r>a  \\
\end{matrix} \right. .
\label{eqn:outsideg2}
\end{equation}
For the Lyot plane fields given by Eqns. \ref{eqn:outsideg} and \ref{eqn:outsideg2}, the on-axis point source is extinguished by a simple circular LS with radius less than $a$. 

\begin{figure}[t!]
\begin{center}
\includegraphics[width=\linewidth,trim =0mm 0mm 0mm 0mm,clip=true]{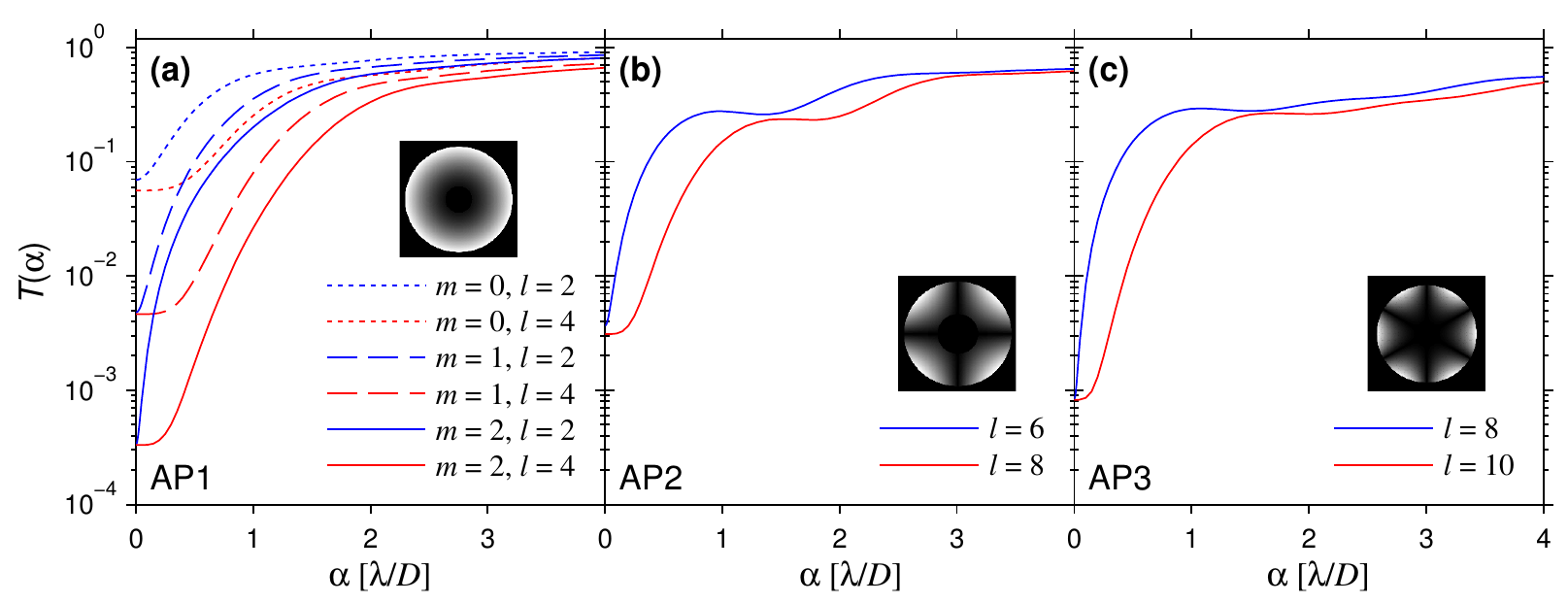}
\end{center}
\caption{ \label{fig:offaxisTransPower_Zernike} 
Transmitted energy for a distant point source at angular displacement $\alpha$ for (a) AP1, (b) AP2, and (c) AP3 with Zernike amplitude apodizers (shown in the inset). } 
\end{figure} 

The three example apodizers shown in Fig. \ref{fig:Zernike} may be particularly useful for improving the starlight suppression provided by a VC with the pupil shapes shown in Fig. \ref{fig:apertures}. Specifically, the Zernike amplitude pupil functions in Fig. \ref{fig:Zernike}(a)-(b) and Fig. \ref{fig:Zernike}(f)-(g) may be matched with AP2 and AP3, respectively, with the dark regions oriented along the radial spiders. The simple annulus (AP1) may also be apodized by Fig. \ref{fig:Zernike}(k)-(l) (or any function $P\left(r,\theta \right)=\left(r/a\right)^{m}e^{im\theta}$). Figure \ref{fig:offaxisTransPower_Zernike} shows the transmitted energy with such apodizers located in PP1 for the apertures Fig. \ref{fig:apertures} and a circular LS. It can be seen that the value of $T\left(0\right)$ is quite low, but the off-axis signal is reduced. In the case of AP1, we find that $T\left(0\right)$ may be reduced by increasing the value of $l$.

There is an additional loss owing to the apodizer transmission given by
\begin{align}
\tau_{n,m}&=\frac{1}{\pi a^2}\int_0^{2\pi}\int_0^a |P(r,\theta)|^2\,r\,dr\,d\theta, \\
&=\frac{1}{\pi}\int_0^{2\pi}\int_0^1 \left[Z_n^m(\rho,\theta)\right]^2\,\rho\,d\rho\,d\theta,\\
&=\left\{ \begin{matrix}
   \left(n+1\right)^{-1} & m=0  \\
   \left(2n+2\right)^{-1} & m\ne0  \\
\end{matrix} \right. .
\end{align}
For the AP1 apodizer $\left(r/a\right)^{m}e^{im\theta}$, the signal is reduced by a factor of $\left(m+1\right)^{-1}$. 

The discovery of an analytical basis for VC apodizers offers a route to complex entrance pupil function optimization. For a given value of $l$, a linear combination of Zernike amplitude apodization functions of the form 
\begin{equation}
P\left(r,\theta\right) = \sum_{n,m}c_{n,m}Z_n^m\left(r/a,\theta\right),\;\;\;\;\;r\le a,
\label{eq:lincomb}
\end{equation}
yield ideal starlight suppression for a circular LS provided $l$ is even and $|l|>\mathrm{max}\left\{N+M\right\}$, where $N$ and $M$ are the maximum values of $n$ and $|m|$, respectively. The coefficients $c_{n,m}$ may be complex constants and each Zernike polynomial may be rotated an arbitrary amount owing to symmetry. Also, the fields inside the LS may be canceled between higher-order opposite-parity counterparts (e.g. $Z_m^m\pm iZ_m^{-m}$). Although most current implementations of the VC have vortex charge $l=2$ or $l=4$, increasing the value of $l$ allows for many more apodization functions to be devised, which may improve performance with very complicated pupil obstructions. In general, the apodizers presented in Fig. \ref{fig:Zernike} may be further optimized for complicated apertures by introducing additional Zernike polynomials with $n+m<|l|$. Furthermore, Zernike polynomials with $n+m>|l|$ may have a negligible effect to the contrast performance if $\left|c_{n,m}\right|^2\ll1$ or are canceled by opposite-parity counterparts. Numerically optimized entrance pupil apodizers based on Zernike amplitude modes will be reported elsewhere.

\subsubsection{Lossless amplitude apodization}

A potential advantage of using an entrance pupil apodizer is that the starlight suppression does not depend of wavelength. Thus, we expect very good performance over large passbands. One drawback is that conventional semi-transparent optics introduce losses that may inhibit detection of dim companions. However, recent developments in pupil remapping for coronagraphy may provide a route to improved transmission performance \cite{Guyon2003,Guyon2005,Pueyo2013,Guyon2014}. The optical design of out-of-plane aspheric optics for lossless (or low loss) Zernike amplitude apodization is underway. 

\subsection{Lyot plane optimization}
Conventional focal-plane coronagraphs only include a binary amplitude LS that has similar shape to the entrance pupil in PP2. Here, we describe phase-only (or complex) Lyot plane masks (LPMs) that improve image plane contrast. Contrast compares the signal from the on-axis point source that appears at a particular location to the signal from an imaged source located at that position. That is, the contrast at angular position $\alpha$ is defined as
\begin{equation}
C\left(\alpha \right)= \frac{\kappa\left(0,\alpha \right)}{\kappa\left(\alpha,\alpha \right)},
\label{eq:contrast}
\end{equation}
where
\begin{equation}
\kappa\left(\alpha_1, \alpha_2\right)= \int_{FP2} \left|h\left(x,y;\alpha_1 \right) \right|^2 \Gamma\left(x,y;\alpha_2 \right) dA,
\end{equation}
$h\left(x,y;\alpha \right)$ is the PSF in FP2 for a point source at angular displacement $\alpha$ and $\Gamma\left(x,y;\alpha \right)$ represents a circular hole centered at $\alpha$ with diameter equal to the full width at half maximum (FWHM) of $\left|h\left(x,y;\alpha \right) \right|^2$. Since the FWHM may vary with $\alpha$, we typically use the FWHM calculated for a point source in the center of the discovery region. 

\begin{figure}[t!]
\centering
\includegraphics[width=0.94\linewidth]{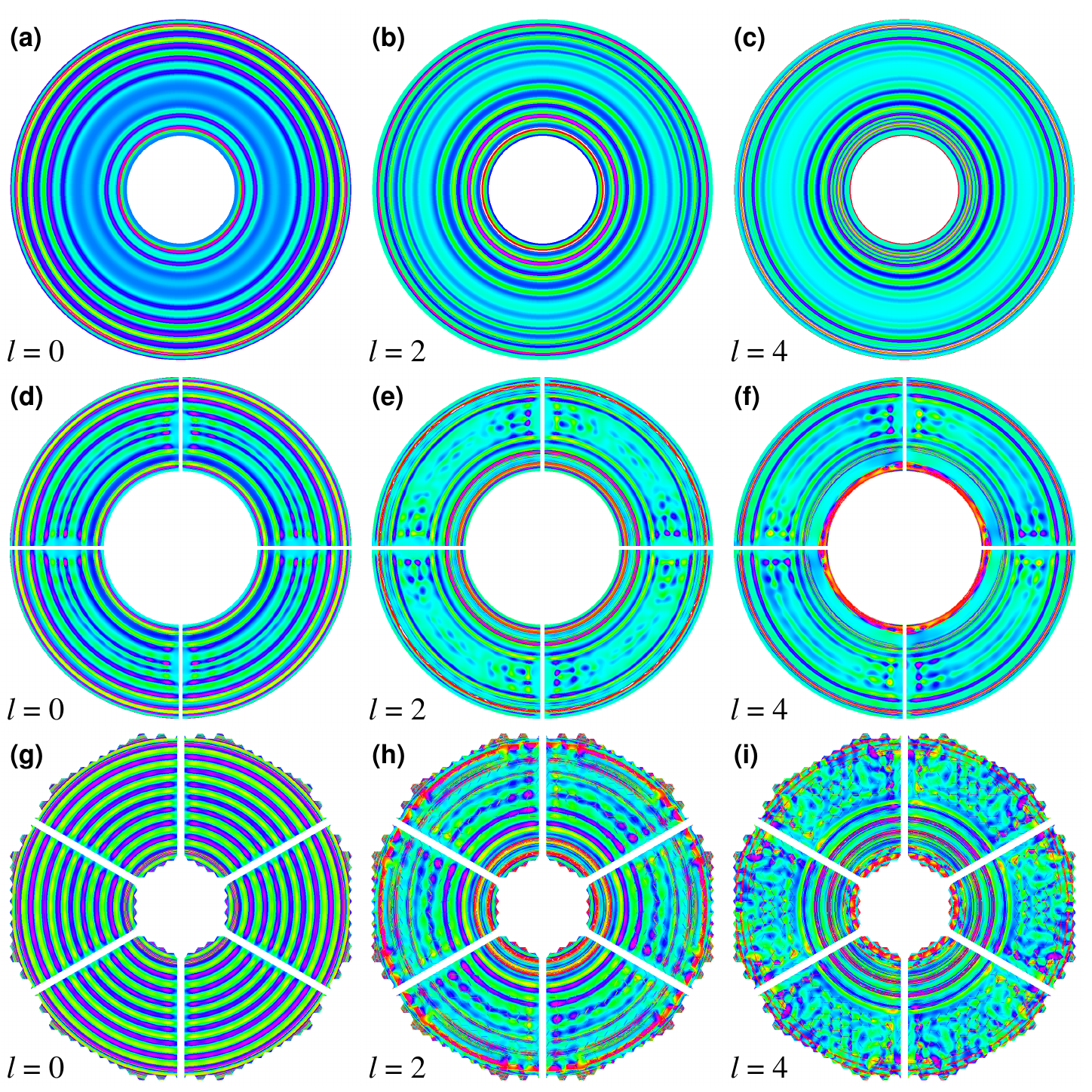}
\includegraphics[trim =53mm -54mm 2mm 11mm,clip=true]{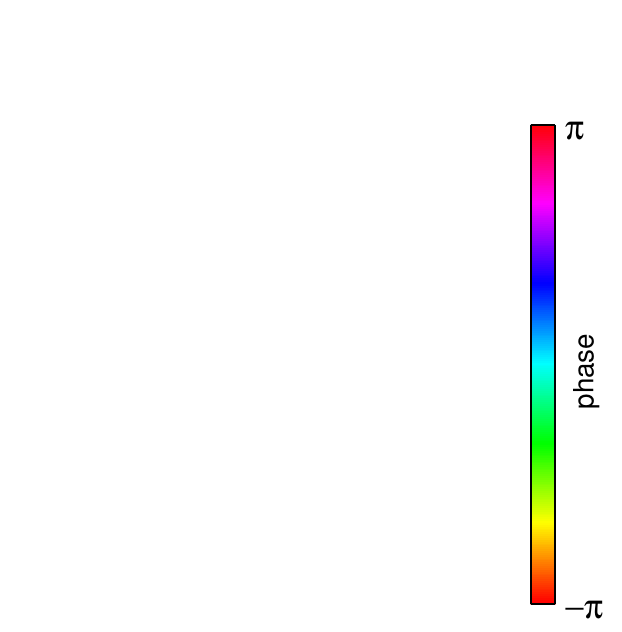}
\caption{ \label{fig:LPMs} 
Phase-only LPMs optimized for (a)-(c) AP1, (d)-(f) AP2, and (g)-(i) AP3 for various values of $l$. The corresponding on-axis PSFs are shown in Fig. \ref{fig:LPMs_PSFs} .}
\end{figure} 

\subsubsection{Phase-only Lyot plane masks}
\begin{figure}[t!]
\centering
\includegraphics[width=0.93\linewidth]{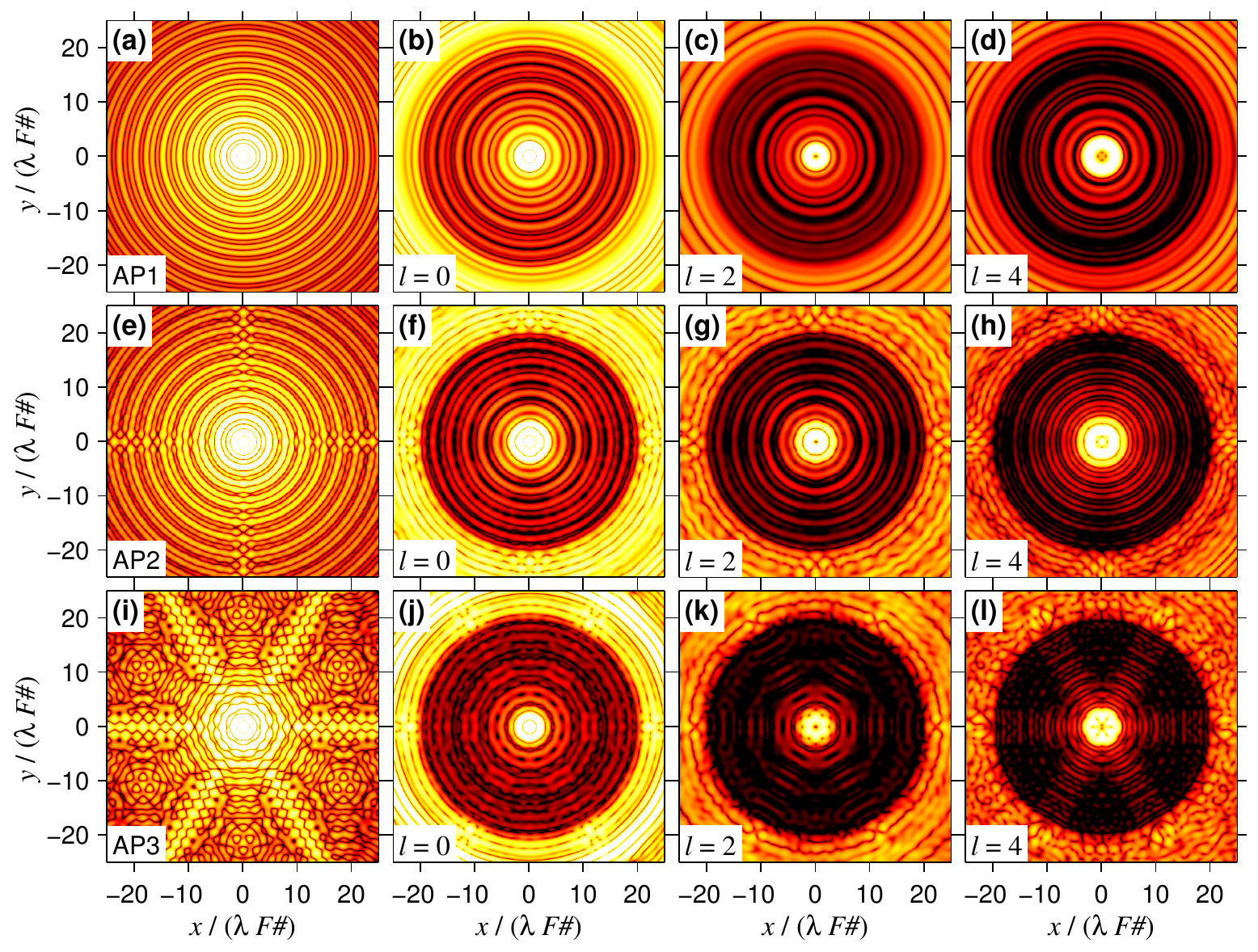}
\includegraphics[trim =53mm -38mm 2mm 11mm,clip=true]{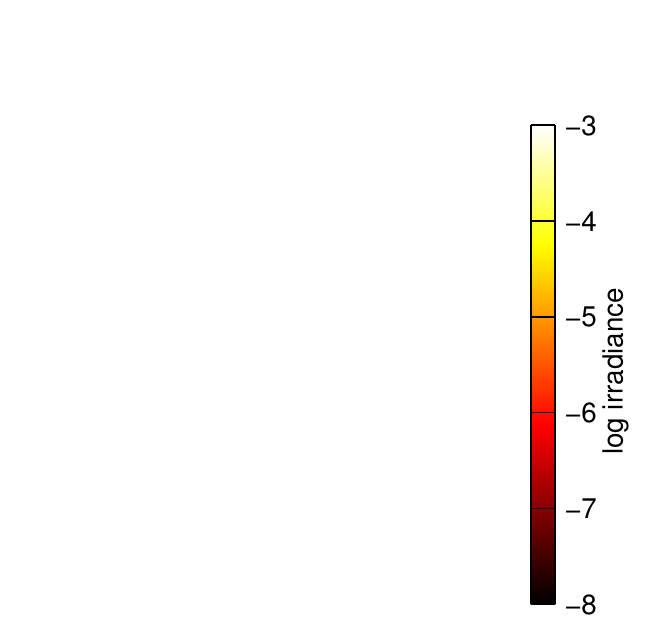}
\caption{ \label{fig:LPMs_PSFs} 
Monochromatic, on-axis, point spread function $\left|h\left(x,y;0 \right) \right|^2$ in log irradiance for (a)-(d) AP1, (e)-(h) AP2, and (i)-(l) AP3. The PSF without masks at PP1, FP1, and PP2 are shown in (a),(e), and (i).}
\end{figure}

We calculate phase-only Lyot plane masks (LPMs) that form a dark hole in the on-axis PSF using a slightly modified algorithm to the one outlined in section \ref{sec:FPC} (also see Ruane et al. 2015, in prep.). For the LPM, the necessary phase in PP2 is calculated to form a dark hole in the on-axis PSF at PP2. The goal is to reduce the contrast between starlight and the off-axis companion in a pre-defined discovery region. Figure \ref{fig:LPMs} shows the phase-only LPMs for the pupils shown in Fig. \ref{fig:apertures}(a)-(c) with charge $l$ vortex phase masks located in FP1. The LPMs for AP1 can be made circularly symmetric, whereas more complicated patterns are needed to optimize the performance with AP2 and AP3. The corresponding on-axis PSFs in each case are shown in Fig. \ref{fig:LPMs_PSFs}. The LPMs suppress the irradiance within a annulus ranging from approximately 4 to 20 $\lambda F\#$. Circular symmetry is forced every 10 iterations for the first 100 iterations and the final phase corrector is calculated using 200 total iterations. This helps encourage a well-behaved solution with limited azimuthal variation. 

\begin{figure}[t!]
\centering
\includegraphics[width=\linewidth]{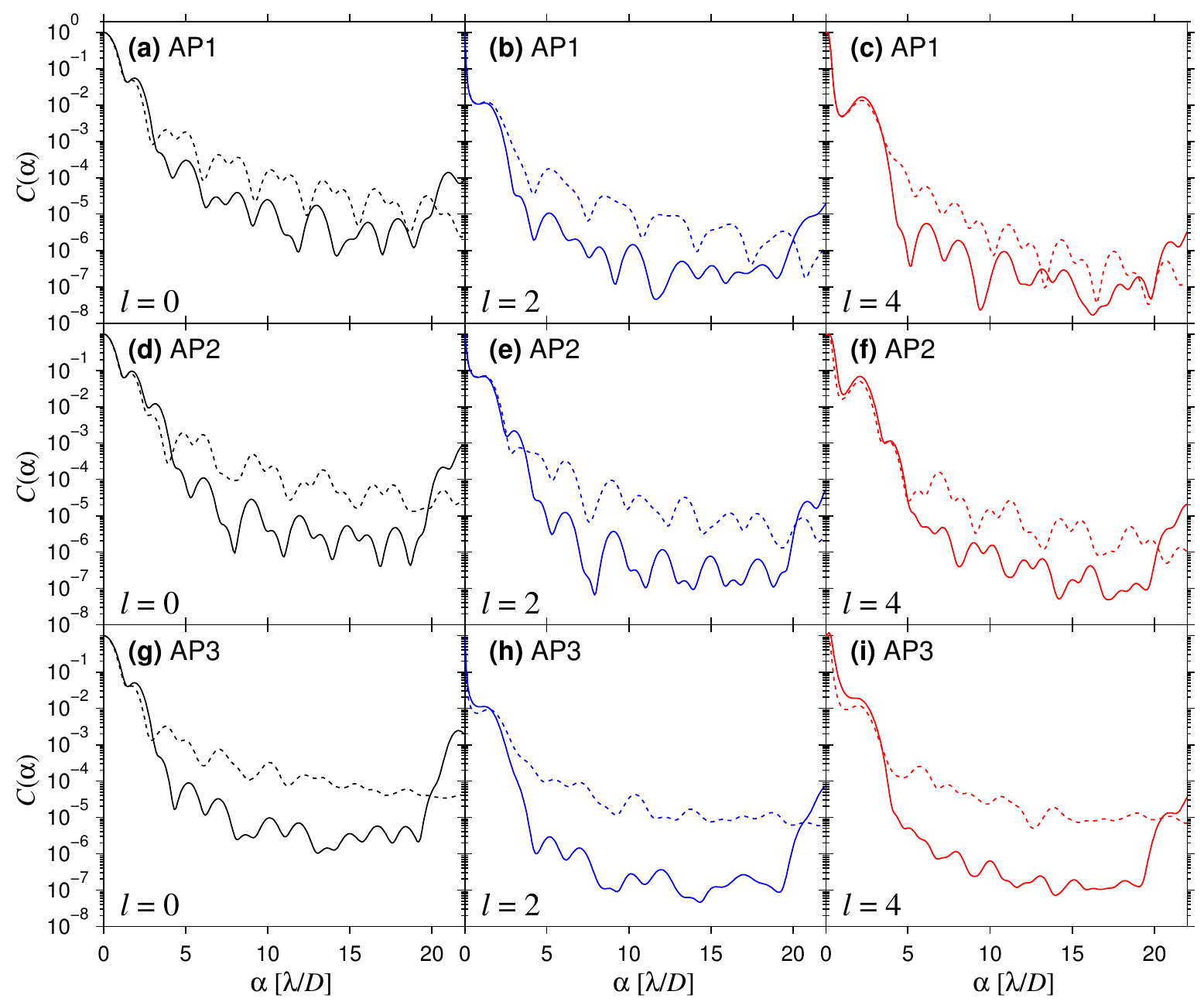}
\caption{ \label{fig:contrast_LPMs} 
Contrast achieved without (dashed lines) and with (solid lines) the LPMs shown in Fig. \ref{fig:LPMs}.}
\end{figure} 

\begin{figure}[t!]
\centering
\includegraphics[trim =1mm -7mm 2mm 0mm,clip=true]{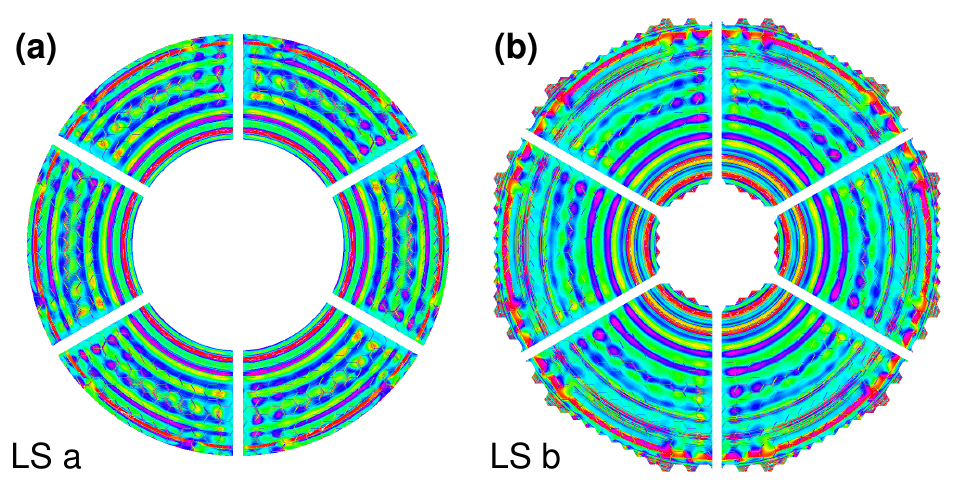}
\includegraphics[trim =53mm -11mm 2mm 8mm,clip=true]{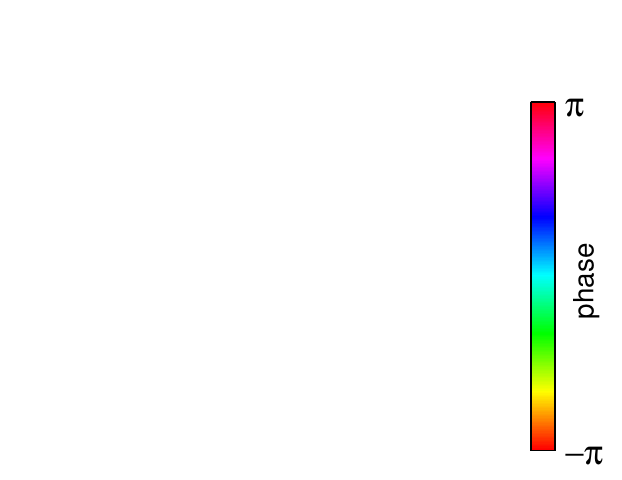}\hfill
\includegraphics[trim =0mm 0mm 0mm 0mm,clip=true]{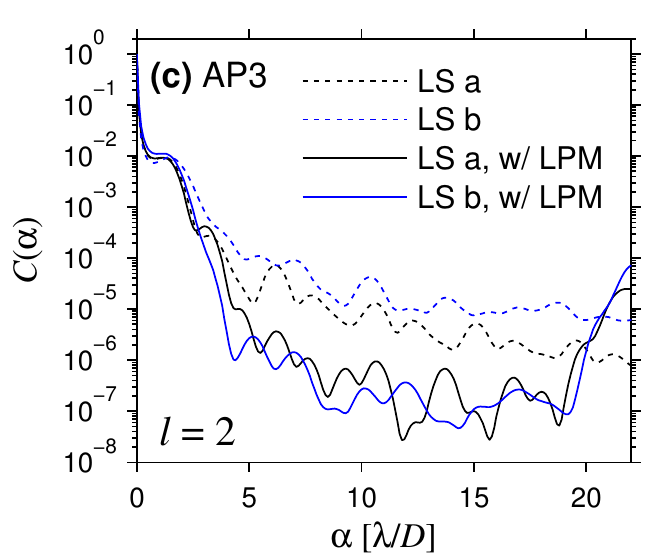}
\caption{ \label{fig:LScompare_LPMs} 
Effect of LS size for an $l=2$ VC on AP3. (a) LPM solution for a significantly downsized LS. (b) LPM solution for an enlarged LS. (c) Contrast achieved with the LPMs shown in (a) and (b).}
\end{figure} 

The introduction of an LPM tends to negatively effect the quality of the off-axis PSF. To take this into account, we report the contrast achieved with the LPMs (see Fig. \ref{fig:contrast_LPMs}). The on-axis PSF is azimuthally averaged and the off-axis PSF is taken to be displaced along the $x$-direction. A reduction in contrast indicates that the starlight is suppressed at $\alpha$ more than the light from the off-axis source and therefore the performance is improved.  

The intrinsic wavelength dependence of the LPM is fully described by the contrast curves shown in Fig. \ref{fig:contrast_LPMs}. The curves are radially blurred in the broadband case according to the scaling of the horizontal axis. Since the optimization region is large, we expect improved performance over a large bandwidth. 

The size and shape of the LS can significantly affect the achieved contrast performance of an LPM. Though using a larger LS allows more starlight to reach FP2, we find that a larger LS with an LPM may lead to better contrast performance in addition to improved transmission for off-axis sources. Figure \ref{fig:LScompare_LPMs} compares solution for the LS in Fig. \ref{fig:apertures}(f) to a larger LS for an $l=2$ VC on AP3. In this case, the large LS leads to improved contrast at several angular intervals, especially within 3--10~$\lambda/D$. Similar relationships are found in the $l=0$ and $l=4$ case. 

We note that an arbitrarily shaped dark hole may be formed in the on-axis PSF. The main effect is that using a smaller region (e.g. a half plane) or moving the inner boundary of the optimization region further from the star allows for deeper contrast to be achieved.  

All simulations described here assume a perfect wavefront entering the coronagraph. The effect of realistic wavefront error on adaptively-corrected high-contrast imaging instruments will be reported elsewhere (Ruane et al. 2015, in prep.).

\subsubsection{Lyot plane complex masks}

Just as we have shown in the case of focal plane correctors in section \ref{sec:FPC}, a complex version of the LPMs may be calculated rather than phase-only LPMs. We generally find that similar performance is achieved with complex solutions. However, adding an amplitude component may be useful for limiting the high-frequency phase variations in the LPMs, but also cause an unwanted loss in signal from off-axis companions. The benefit of using complex LPMs will be further investigated in future work. 

\section{CONCLUSIONS AND FUTURE OUTLOOK} \label{sec:concl} 
We have described three ways to improve the performance of vortex coronagraphs on telescopes with complicated apertures. Optical elements that vary the complex amplitude and phase may be introduced in the entrance pupil, focal plane, and/or Lyot plane to suppress the starlight while maintaining image quality and sensitivity to dim off-axis companions. We have presented both analytical and numerical approaches for designing pupil and focal plane masks. Future work will investigate using the presented methods to jointly optimize all three optics for contrast and throughput as well as reduced chromatic dependence and sensitivity to aberrations. Using these methods, it is possible to tailor a three-plane coronagraph for arbitrarily complicated telescopes pupils, such as heavily obstructed or segmented apertures. 

\acknowledgments     
This work has benefited from computing assistance from Carlos Gomez Gonzalez (Universit\'{e} de Li\`{e}ge, Belgium) as well as fruitful discussions with Prof. Jean Surdej (Universit\'{e} de Li\`{e}ge, Belgium) and Prof. Matt Kenworthy (Leiden Observatory, Netherlands). G. J. Ruane was supported by Wallonie-Bruxelles International's (Belgium) Scholarship for Excellence and the U.S. National Science Foundation under Grant No. ECCS-1309517. The research leading to these results has received funding from the European Research Council under the European Union's Seventh Framework Programme (ERC Grant Agreement n. 337569) and from the French Community of Belgium through an ARC grant for Concerted Research Action. 

\bibliographystyle{spiebib} 
\bibliography{Ruane_SPIE_OP2015} 

\end{document}